**Charge density wave instability and pressure-induced superconductivity in bulk 1*T*-NbS$_2$**

Wei Wang,[1] Bosen Wang,[2] Zhibin Gao,[3] Gang Tang,[4] Wen Lei,[1] Xiaojun Zheng,[1] Huan Li,[1] Xing Ming,[1*] and Carmine Autieri[5†]

*1. College of Science, Guilin University of Technology, Guilin 541004, China*

*2. Beijing National Laboratory for Condensed Matter Physics and Institute of Physics, Chinese Academy of Sciences, Beijing 100190, China*

*3. Department of Physics, National University of Singapore, Singapore 117551, Republic of Singapore*

*4. Theoretical Materials Physics, Q-MAT, CESAM, University of Liège, B-4000 Liège, Belgium*

*5. International Research Centre MagTop, Institute of Physics, Polish Academy of Sciences, Aleja Lotników 32/46, PL-02668 Warsaw, Poland*

## ABSTRACT

Charge-density-wave (CDW) instability and pressure-induced superconductivity in bulk 1*T*-NbS$_2$ are predicted theoretically by first-principles calculations. We reveal a CDW instability towards the formation of a stable commensurate CDW order, resulting in a $\sqrt{13}\times\sqrt{13}$ structural reconstruction featured with star-of-David clusters. The CDW phase exhibits one-dimensional metallic behavior with in-plane flat-band characteristics, and coexists with an orbital-density-wave order predominantly contributed by $4d_{z^2-r^2}$ orbital from the inner Nb atoms of the star-of-David cluster. By doubling the cell of the CCDW phase along the layer stacking direction, a metal-insulator transition may be realized in the CDW phase in case the interlayer antiferromagnetic ordering and Coulomb correlation effect have been considered simultaneously. Bare electron susceptibility, phonon linewidth and electron-phonon coupling calculations suggest that the CDW instability is driven by softened phonon modes due to the strong electron-phonon coupling interactions. CDW order can be suppressed by pressure, concomitant with appearance of superconductivity. Our theoretical predictions call for experimental investigations to further clarify the transport and magnetic properties of 1*T*-NbS$_2$. Furthermore, it would also be very interesting to explore the possibility to realize the CDW order coexisting with the superconductivity in bulk 1*T*-NbS$_2$.

[*] Email: mingxing@glut.edu.cn (Xing Ming)

[†] Email: autieri@magtop.ifpan.edu.pl (Carmine Autieri)

## I. INTRODUCTION

Quasi two-dimensional (2D) layered transition metal dichalcogenides (TMDCs) have attracted extensive attention in recent years. TMDC materials often crystallize in 1*T* or 2*H* polymorph with octahedral or trigonal prismatic coordination [1]. Especially, many group-**V** TMDC materials $MX_2$ (M denotes transition metals Nb or Ta, and X stands for chalcogen elements Se or S) exhibit interesting phenomena and properties, such as charge density wave (CDW) order, superconductivity, metal-insulator transitions, magnetic ordering and Mott physics [2-13]. However, even though possessing the same $MX_2$ chemical composition, different polytypes of $MX_2$ exhibit completely different physical properties. For example, 1*T* polymorph of $TaS_2$ and $TaSe_2$ show complex CDW phase diagrams including incommensurate, nearly commensurate, and commensurate CDW (CCDW) phases with prototypical $\sqrt{13} \times \sqrt{13}$ star-of-David structural reconstructions upon decreasing temperature [2,13,14]. Electronic structure transition arising from Mott-Hubbard physics is observed in the CCDW phase of 1*T*-$TaS_2$ [13], whereas the Mott insulating phase does not appear in bulk 1*T*-$TaSe_2$ even in the CCDW state [14]. In addition, the CDW instability in 1*T* polymorph of $TaS_2$ and $TaSe_2$ can be suppressed under high pressure, then the associated superconductivity emerges [13,15-19]. By contrast, the CDW order and superconductivity can coexist in the 2*H*-$MX_2$, exhibiting complicated competitive or cooperative relations [4,20-22]. Furthermore, the CDW phase transitions display a 3 × 3 periodicity in the bulk form of 2*H*-$MX_2$ [4,11,12]. However, previous researches indicate that 2*H*-$NbS_2$ does not exhibit CDW order due to anharmonic suppression [23,24]. Recent experiments observe very weak superlattice peaks in 2*H*-$NbS_2$ corresponding to a commensurate $\sqrt{13} \times \sqrt{13}$ periodic lattice distortion, identical to the cases in 1*T* polymorph of $TaS_2$ and $TaSe_2$, which may be stem from a local 1*T*-like environment in the 2*H* crystal arising from stacking faults [25].

In addition to the significant roles of the different polymorphs, the dimensionality also plays a profound impact on the transport properties, CDW order and superconductivity of $MX_2$. The CDW phase of the monolayer 1*T*-$TaSe_2$ is identified to be a Mott insulator with unusual orbital texture, but the energy gap reduces significantly for the bilayer, while the trilayer shows semimetallic behavior and the bulk is a one-dimensional (1D) metal [14,26,27]. Contrary to the absence of CDW order in the bulk phase, a 3 × 3 CDW order has been observed down to single-layer limit of 2*H*-$NbS_2$, and it is so fragile that can be disturbed by small compressive strains [28-30]. Strain-induced phase separation between triangular and stripe phases in charge-ordered 1$H$-$NbSe_2$ monolayer have been explored by *ab initio* calculations [31]. Compared to the bulk 2$H$ polymorph of $NbSe_2$ and $TaSe_2$, the CDW instability can be strongly enhanced in their monolayer limit [32,33]. By contrast, the CDW order of bulk 2$H$-$TaS_2$ vanishes in the 2D limit, accompanied with a substantial enhancement of the superconductivity, indicating a competition between CDW order and superconductivity down to the monolayer limit [34,35]. Although it is difficult to synthesize bulk 1$T$ polymorph of $NbSe_2$ and $NbS_2$, few-layer thin film and monolayer have been prepared successfully [36-39]. Similar to other 1$T$ polytype CDW materials, the monolayer 1$T$-$NbSe_2$ is found to be a Mott insulator concomitant with the typical $\sqrt{13}\times\sqrt{13}$ CDW order [38-41], and the monolayer 1$T$-$NbS_2$ is predicted theoretically to undergo a $\sqrt{13}\times\sqrt{13}$ star-of-David structural reconstruction stabilized in spin 1/2 magnetic insulating state [42].

The purpose of present paper is to further explore the existence of CDW instability in bulk 1$T$-$NbS_2$ by first-principles calculations. We firstly discover that the undistorted high-symmetry 1$T$-$NbS_2$ is unstable with softening phonon modes, and undergoes a $\sqrt{13}\times\sqrt{13}$ structural reconstruction to form stable CCDW phase. Electronic structure calculations indicate that orbital density wave (ODW) coexists with the CDW order in the CCDW phase featured with flat-band and 1D metallic characteristics. Once an interlayer antiferromagnetic (AFM) ordering and on-site Coulomb repulsion interactions are taken into account simultaneously, a metal-insulator transition maybe observed by doubling the cell of the CCDW phase. By analyzing the Fermi surface nesting function and calculating the electron-phonon coupling (EPC) constants, we propose that the CDW instability mainly arises from the strong electron-phonon interactions. The CDW order can be suppressed under high pressure, accompanied by the emergence of the superconductivity in the compressed phase. Further experimental and theoretical investigations on the structure and transport behavior of 1$T$-$NbS_2$ under pressure are deserved to better address the intricate competitive or cooperative relations between CDW order and superconductivity in this material.

## II. COMPUTATIONAL DETAILS

The density functional theory (DFT) calculations were performed using the *QUANTUM ESPRESSO* package [43] with generalized gradient approximation (GGA) according to the Perdew-Burke-Ernzerhof (PBE) functional [44]. The ultrasoft pseudopotentials were used to describe the interactions between electrons and ionic cores [45]. The energy cutoff of 45 Ry (540 Ry for the charge density) was chosen for our calculations. Considering the 2D layered structure of the bulk 1*T*-$NbS_2$, we performed geometrical optimization with van der Waals (vdW) corrections of DFT-D3 to obtain accurate lattice constants consistent with experiments [46]. The Brillouin zone (BZ) was segmented by a 16 × 16 × 8 Monkhorst and Pack (MP) grid for undistorted 1*T*-$NbS_2$, while a 4 × 4 × 8 MP grid was used for the CCDW structure [47]. The total energy and electron charge density calculations were performed by Gaussian smearing method with a smearing parameter σ of 0.01 Ry. Using density functional perturbation theory, phonon dispersion curves and EPC were calculated on a 8 × 8 × 4 $\boldsymbol{q}$ grid, and a denser 32 × 32 × 16 $k$ mesh were used for electron-phonon calculation for undistorted structure [48]. Electronic structures and density of states (DOS) are calculated by Vienna *ab initio* simulation package (VASP) [49,50] within GGA approach [44] and projector-augmented wave (PAW) potentials [51], where energy cutoff of 520 eV was used. To more accurately describe the electronic correlations, an on-site Coulomb interaction $U$ with a value of 2.95 eV was considered for the Nb 4$d$ shell to calculate the electronic structure [40]. The Fermi surfaces are visualized using the *FermiSurfer* code [52].

## III. RESULTS AND DISCUSSIONS

### A. Undistorted high-symmetry structure of 1*T*-$NbS_2$

As illustrated in **Fig. 1(a)**, bulk 1*T*-$NbS_2$ displays a layered structure with space group $P\bar{3}m1$, its corresponding BZ and high-symmetry points are illustrated in **Fig. 1(b)**. Each Nb atom is surrounded by six nearest-neighbor S atoms with octahedral coordination, where the adjacent layers are held together by vdW forces. 1*T*-$NbS_2$ film has been synthesized successfully by atmospheric pressure chemical vapor deposition at 600 °C with lattice parameters $a$ = 3.4206 Å and $c$ = 5.9381 Å [36]. Our optimized lattice parameters ($a$ = 3.365 Å and $c$ = 5.954 Å) are in good agreement with the experimental data.

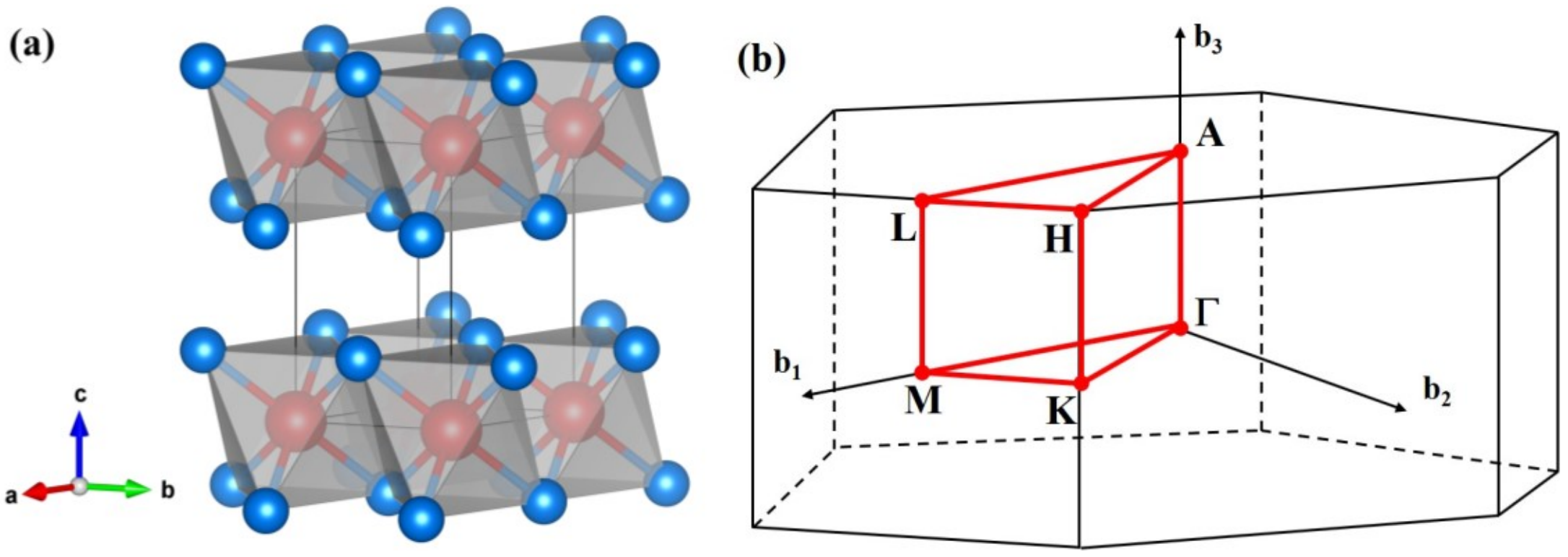

**FIG. 1.** (a) Crystal structure of the $1T$-$NbS_2$ bulk and (b) its corresponding BZ. Nb and S atoms are drawn by red and blue balls in (a), respectively. The irreducible BZ and high-symmetry points are indicated in (b), where **Γ** point is at the zone center.

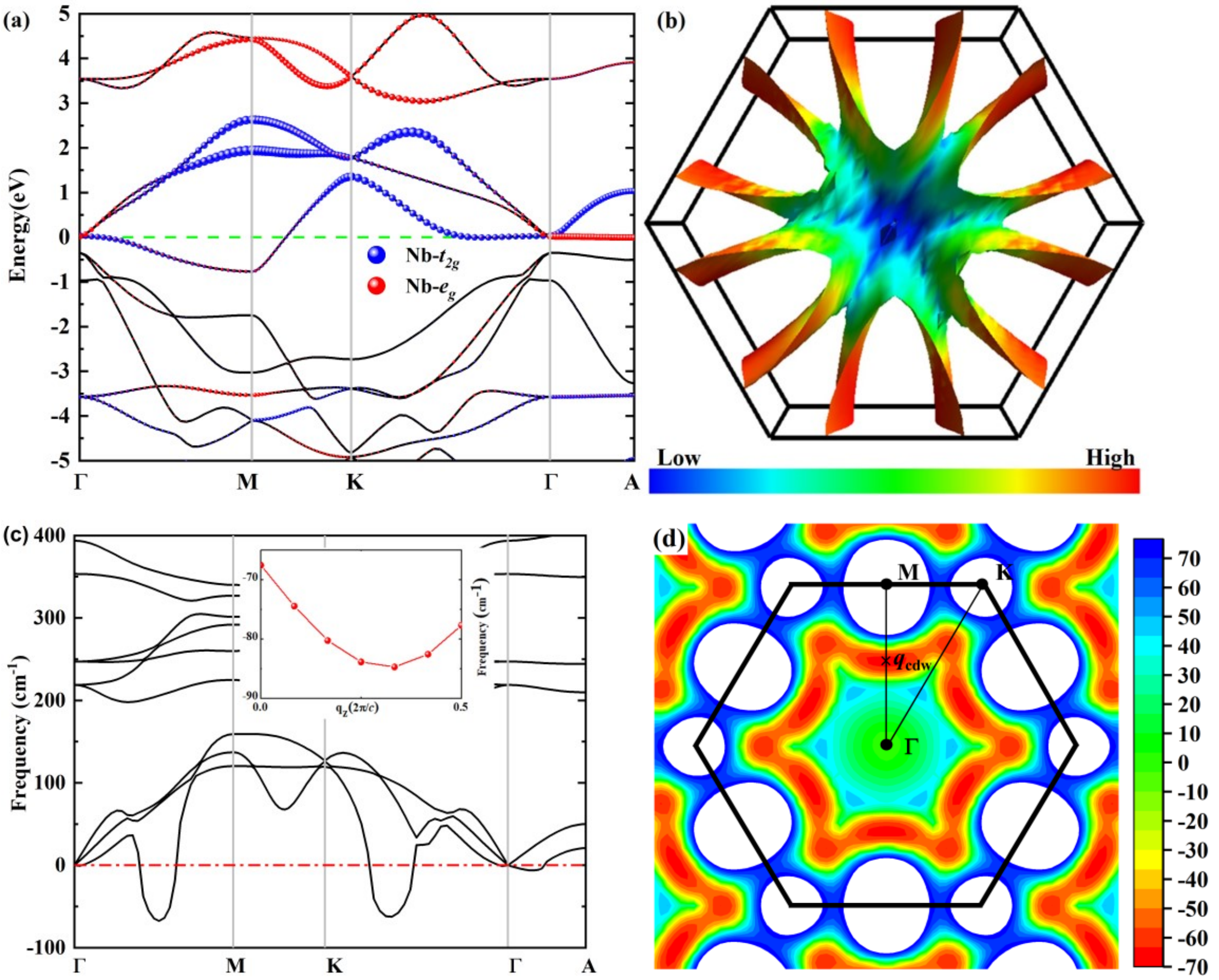

**FIG. 2.** Electronic structure and phonon spectra of the bulk $1T$-$NbS_2$ in the high-symmetry phase: (a) Band structure, (b) Fermi surface, (c) Phonon dispersion curves and (d) Distribution of phonon frequencies at $q_z = 0$ plane. For the band structure in (a), the dashed line in green color corresponds to the Fermi level ($E_F$), and the contributions of Nb $t_{2g}$ ($e_g$) states are proportional to the size of the blue (red) balls. The color bar of the Fermi surface in (c) denotes the Fermi velocity. Imaginary

frequencies in (c) represent unstable modes, and the inset shows the $q_z$ dependence of the unstable acoustic branch. The white regions in (d) indicate phonon frequencies far above 70 $cm^{-1}$.

To obtain fundamental features of the electronic structure of the high-symmetry phase, we firstly calculate the band structure and the Fermi surface of the bulk 1*T*-$NbS_2$. As shown in **Fig. 2**, the electronic structure exhibits similar features with other 1*T* polymorph $MX_2$ materials [53-55]. The bands around Fermi energy ($E_F$) are dominantly contributed by Nb-4*d* orbitals, which are separated from the S-3*p* states by a gap of about 0.5 eV below the $E_F$ [**Fig. 2(a)**]. Due to the quasi 2D nature of the layered structure, the two bands around the gap are nearly flat along the **Γ-A** direction. Consistent with the octahedral crystal-field splitting, the five 4*d* orbitals are split into triply degenerate $t_{2g}$ ($d_{x^2-y^2}$, $d_{z^2-r^2}$, $d_{xy}$) and doubly degenerate $e_g$ ($d_{yz}$, $d_{xz}$) orbitals [40,42]. The bandwidths of the $t_{2g}$ (~3.5 eV) and $e_g$ (~3 eV) bands are rather large, which implying moderate electron-electron interactions [13]. Especially, the $t_{2g}$ bands crossing Fermi level are filled with one 4*d* electron, forming 2D electron pockets around the **M** point. Similar to other 1*T*-$MX_2$, the Fermi surface features with six-fold petal structural sheets and weak dispersion along the $k_z$ direction [**Fig. 2(b)**], clearly reflects the quasi 2D nature of the electronic states, which is consistent with the layered structure of 1*T*-$NbS_2$ [53-55].

Generally, it is an effective method to predict the CDW instability by first-principles phonon calculation [15,16,41,56,57]. The phonon spectra of the bulk 1*T*-$NbS_2$ are presented in **Fig. 2(c)**. The phonon instability locating at a distinctive position is believed to be directly related to the CDW distortion, which exhibiting softened phonon modes with imaginary frequency at the CDW vector ($\boldsymbol{q}_{\mathrm{CDW}}$). The imaginary phonon frequency at $\boldsymbol{q}_{\mathrm{CDW}}$ signifies structure reconstructions with a superlattice vector of $\boldsymbol{q}_{\mathrm{CDW}}$ [2]. Furthermore, the distributions of phonon frequencies at the plane of $q_z = 0$ in reciprocal space [**Fig. 2(d)**] confirm that the center of CDW instability is located definitely between **Γ** and **M** at $\boldsymbol{q}_{\mathrm{CDW}}$. The CDW instabilities of group-**V** dichalcogenides $MX_2$ are often characterized by $\boldsymbol{q}_{\mathrm{CDW}} \approx 0.25\text{-}0.33 \times \mathbf{\Gamma M}$, depending on different materials [2,41,58]. In bulk 1*T*-$NbS_2$, the maximally instable acoustic softened mode is located at $\boldsymbol{q}_{\mathrm{CDW}} \approx 0.267 \times \mathbf{\Gamma M}$, which is very close to the ordering vector of $\boldsymbol{q}_{\boldsymbol{CCDW}} = 1/\sqrt{13} \times \mathbf{\Gamma M}$, the nearest vector along **Γ-M** compatible with a $\sqrt{13} \times \sqrt{13}$ distortion [16]. This instability persists at all values of $q_z$ as shown in the inset of **Fig. 2(c)**. Moreover, similar to 1*T* polymorph of $NbSe_2$ and $TaTe_2$, there is a larger area of instability expanding to the **Γ-K** line [41,57]. In the **Γ-A** direction, the flatness of the optical branch implies the almost 2D dispersion, which is in line with the quasi 2D crystal structure of bulk 1*T*-$NbS_2$. We note that the CDW order is absent in 2*H*-$NbS_2$ due to anharmonic suppression, which has been demonstrated by the experimental phonon spectrum and fully anharmonic phonon dispersion curves calculation [23,24]. However, most of first-principles investigations on the CDW order of TMDCs have not considered the anharmonic effects, but are still in good agreement with experiments [40,53].

### B. Charge density wave phase and orbital density wave of 1*T*-$NbS_2$

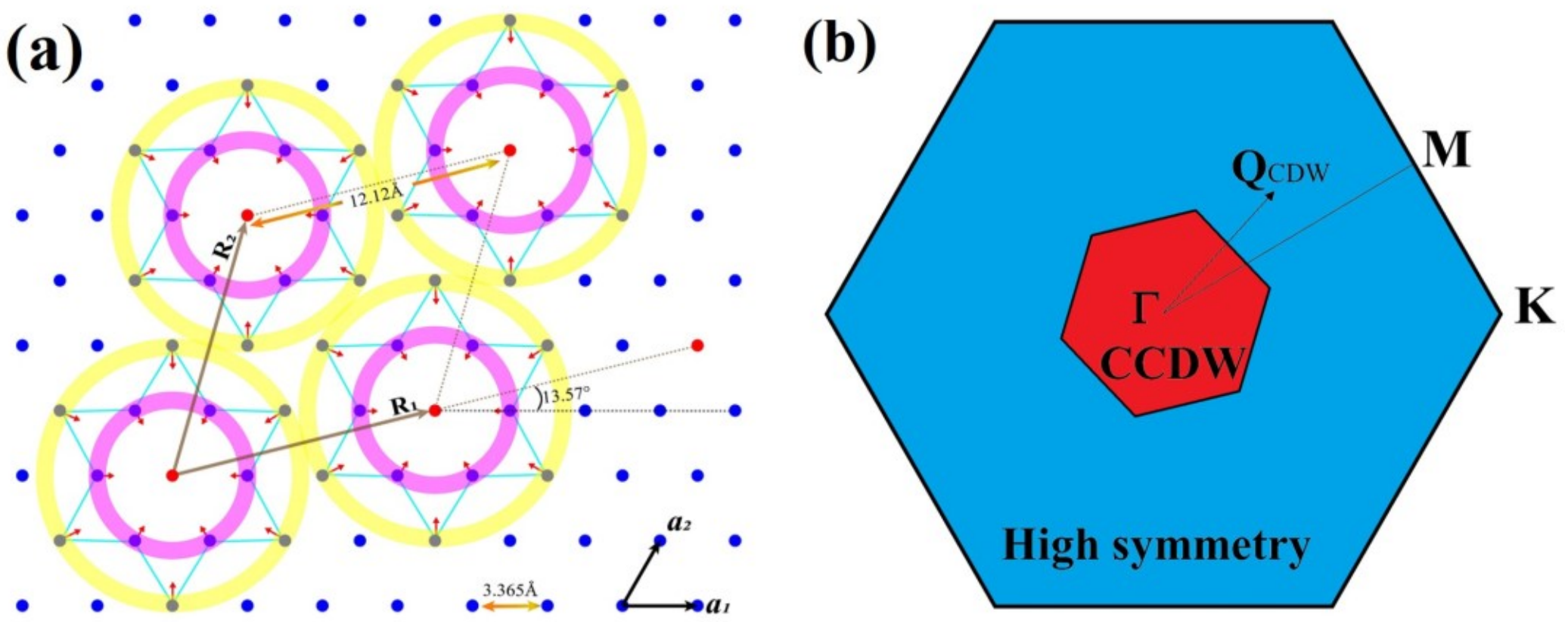

**FIG. 3.** (a) Schematic illustration of the distortion mode in the Nb-atoms layer. The star-of-David clusters are shown in Cyan stars. The red arrows illustrate the distortion directions of the Nb atoms. Purple and yellow circles depict the first-ring and second-ring of the Nb atoms in the star-of-David cluster, respectively. $\mathbf{R}_1$ and $\mathbf{R}_2$ are lattice vectors of CCDW phase, whereas $\boldsymbol{a}_1$ and $\boldsymbol{a}_2$ are lattice vectors of the high-symmetry phase ($\mathbf{R}_1=3\boldsymbol{a}_1+\boldsymbol{a}_2$, $\mathbf{R}_2=-\boldsymbol{a}_1+4\boldsymbol{a}_2$). $a^* = 12.12$ Å is the lattice constant of CCDW phase, while $a = 3.365$ Å is the in-plane lattice constant of the high-symmetry phase. (b) In-plane BZ reconstruction for the CCDW transition. The big blue region represents the first BZ of the high-symmetry phase. The small red hexagon stands for BZ of the CCDW phase. The BZ of the CCDW phase is rotated by 13.57◦ with respect to that of the high-symmetry phase.

After observing the phonon instability for the undistorted high-symmetry phase, we recognize that the high-symmetry structure will undergo a $\sqrt{13}\times\sqrt{13}$ structural reconstruction and transform to a stable CCDW phase with star-of-David clusters. Assuming a nonmagnetic ground state, we perform structural relaxation and obtain a CCDW structure depicted in Fig. 3(a). The $\sqrt{13}\times\sqrt{13}$ superlattice is consisted of Nb clusters with two shells of six Nb atoms and a central thirteenth Nb atom coordinated by five kinds of S atoms (detailed atomic positions and structural mode are presented in **Table S1** and **Fig. S1** in the Supplemental Material [59]). The lattice constants of the CDW structure are $a^* = 12.12$ Å and $c^* = 5.97$ Å with space group $P\bar{3}$. As shown in **Fig. 3(a)**, the central Nb atoms keep the same positions relative to the high-symmetry phase, while the peripheral Nb atoms slightly contract inward toward the central Nb atoms by 5.7% (inner one) and 4.3% (outer one), forming 13-atom star-of-David clusters in the Nb-atoms plane, similar to the bulk 1*T*-$TaS_2$ [60]. As shown in **Fig. S1** of the Supplemental Material [59], inside of the $\sqrt{13}\times\sqrt{13}$ star-of-David clusters, the distances between two nearest-neighbor Nb atoms are shorter than those in the undistorted 1*T*-$NbS_2$, leading to smaller Nb-S-Nb angles with respect to those of the undistorted phase. Conversely, on the boundary of two star-of-David clusters, the distances of two Nb atoms are longer and the Nb-S-Nb angles are larger. Furthermore, the S atoms outside the plane are affected by the in-plane displacements of Nb atoms, resulting in an out-of-plane buckling, which plays an important role in the orbital hybridization [53]. The in-plane Nb atoms' distortion mode and the corresponding BZ reconstruction are presented in **Fig. 3**. Due to the lattice distortion, the lattice vectors of the CCDW phase rotate by 13.57° with respect to the high-symmetry phase in real space. Accordingly, as shown in **Fig. 3(b)**, the BZ of the CCDW phase also rotates by 13.57°, resulting in an in-plane $\sqrt{13}\times\sqrt{13}\mathrm{R} = 13.57^\circ$ periodic lattice distortion [60]. As shown in **Fig. S2** of the Supplemental Material [59], all frequencies of the phonon dispersion for the $\sqrt{13}\times\sqrt{13}$ CCDW phase are positive, revealing that the CCDW phase is dynamically stable, and the nonmagnetic state of the CCDW phase is found to be more stable by 20 meV/$NbS_2$ with respect to the high-symmetry phase.

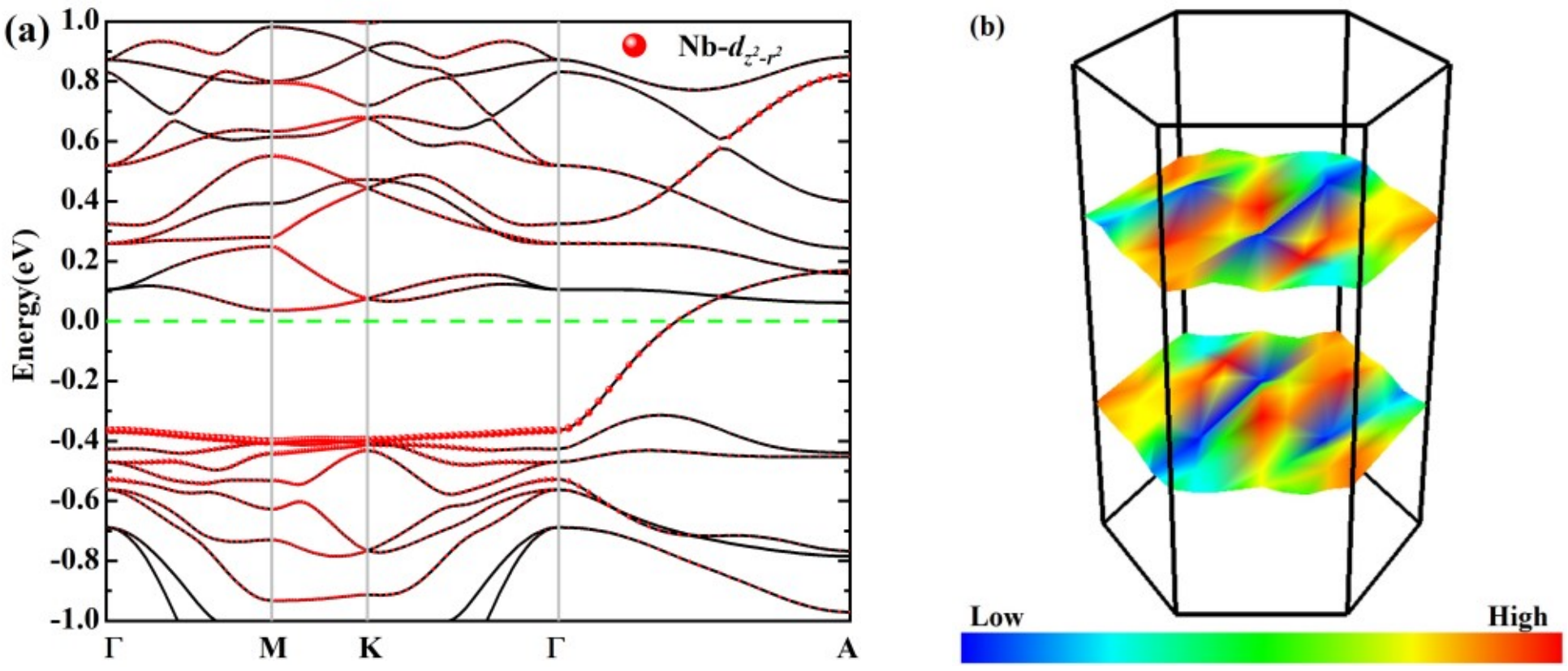

(c)

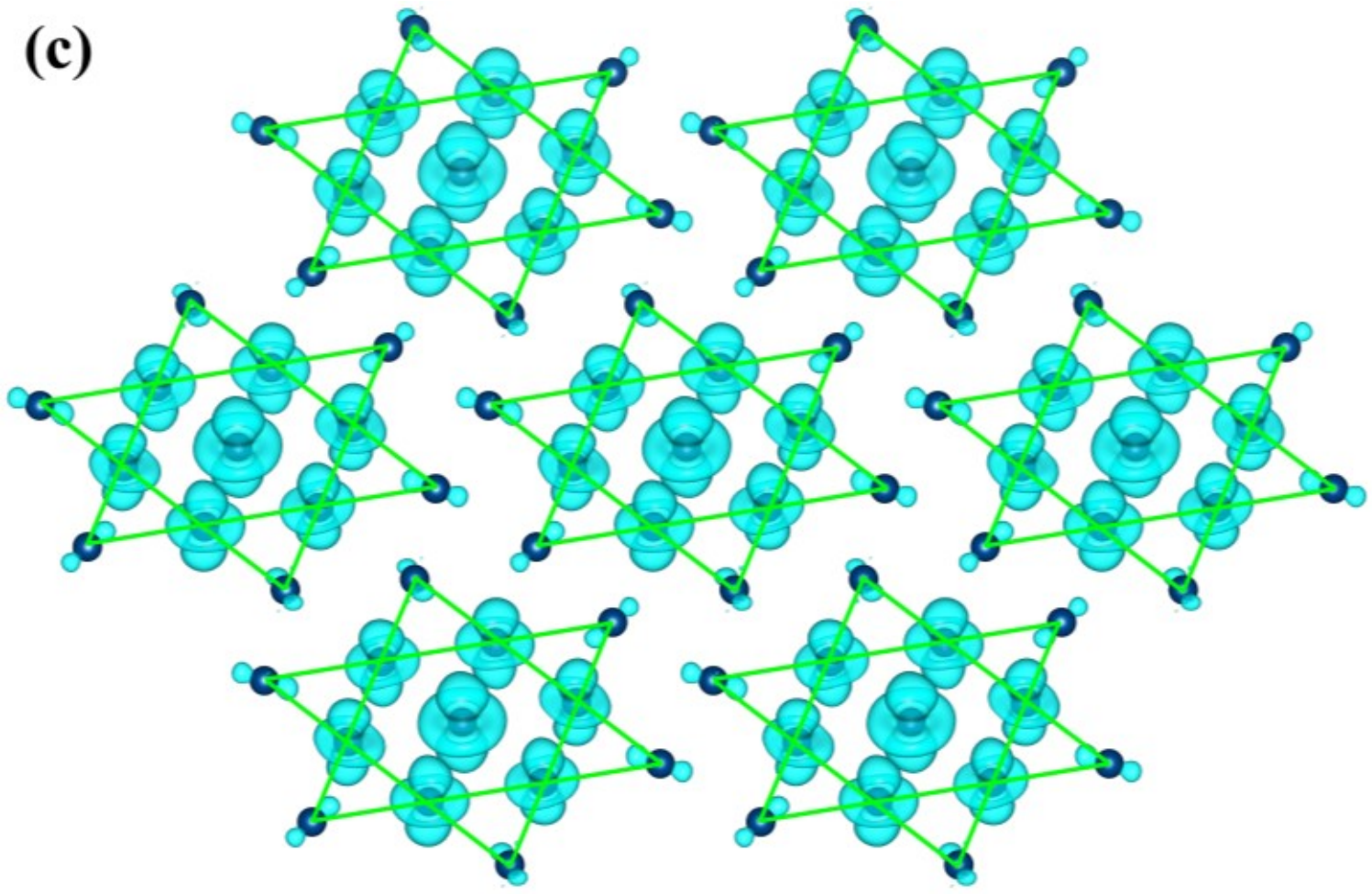

**FIG. 4.** Electronic structures of the bulk 1*T*-$NbS_2$ in the CCDW phase. (a) Band structure calculated within GGA, the red balls represent orbital contributions with the $d_{z^2-r^2}$ character. (b) Fermi surface of the CCDW phase, where the color bar denotes the Fermi velocity. (c) The charge density distribution of the uppermost occupied states in real space, the isosurface value is 0.002 electrons/ Å$^3$.

The structural distortion of the CCDW phase can always lead to electronic structure variations. Compared with the band structures of the high-symmetry phase, we find significant changes around $E_F$ in the CCDW phase. As shown in **Fig. 4(a)**, the band structures show a remarkable in-plane gap of about 0.40 eV, which displays a very weak dispersion (flat-band characteristic) along the in-plane **Γ-M-K-Γ** direction but a strong out-of-plane dispersion along the **Γ-A** direction. In contrast to the quasi-2D flower-shaped Fermi surface of the high-symmetry undistorted phase, the CCDW phase exhibits a pair of quasi-1D Fermi sheets [**Fig. 4(b)**]. Such a feature of the electronic structure can be described as 1D metal, which is in agreement with previous first-principles results for other bulk 1*T* polymorph such as $TaS_2$ and $TaSe_2$ [12,16,53]. The in-plane flat-band implies very heavy effective mass and a small in-plane Fermi velocity. Therefore, the electrons are difficult to hop in the plane. The CCDW phase of the 1*T*-$NbS_2$ can be viewed as an in-plane semiconductor [6]. The out-of-plane metallic transport property and the resistivity anisotropy have been revealed by highly accurate in-plane and out-of-plane electrical resistivities measurements for bulk 1*T*-$TaS_2$ [61]. Therefore, the transport property of bulk 1*T*-$NbS_2$ at low temperatures deserves further experimental measurement.

Furthermore, an interesting orbital texture within the *ab*-plane emerges, which is intertwined with the CCDW [6,26,53]. The projected band structures and real-space charge density distributions illustrate that the highest occupied band arises dominantly from the Nb-4$d$ orbitals with the $d_{z^2-r^2}$ character as shown in **Fig. S3** of the Supplemental Material [59] and **Fig. 4(c)**. Especially, such an orbital configuration is derived mainly from the inner Nb atoms in the star-of-David clusters. The contribution from the Nb atoms at the edge of the clusters is negligible. Therefore, an orbital density wave (ODW) order coexists with the CDW order, which is similar to the in-plane orbital texture of the 1$T$ polymorph of $TaS_2$ and $TaSe_2$ [6,26,53]. The ODW and the CCDW coexist simultaneously with an identical spatial symmetry. The ODW indicates very little in-plane overlap between orbitals centered on neighboring clusters, while charge can only hop significantly along the layer stacking direction ($c$ axis), in agreement with the in-plane flat-band characteristics and out-of-plane quasi-1D metallic nature of the electronic structure. The ODW discussed here is different from the orbital ordering studied extensively in strongly correlated electron systems [62]. The conventional orbital ordering is stemmed mainly from the cooperative Jahn-Teller distortions or Kugel-Khomskii superexchange interactions [63,64], whereas the CCDW and the ODW in 1$T$ polymorph of $MX_2$ are mainly attributed to electron-phonon coupling interactions [6,53].

### C. Possibility of metal-insulator transition in the CCDW phase

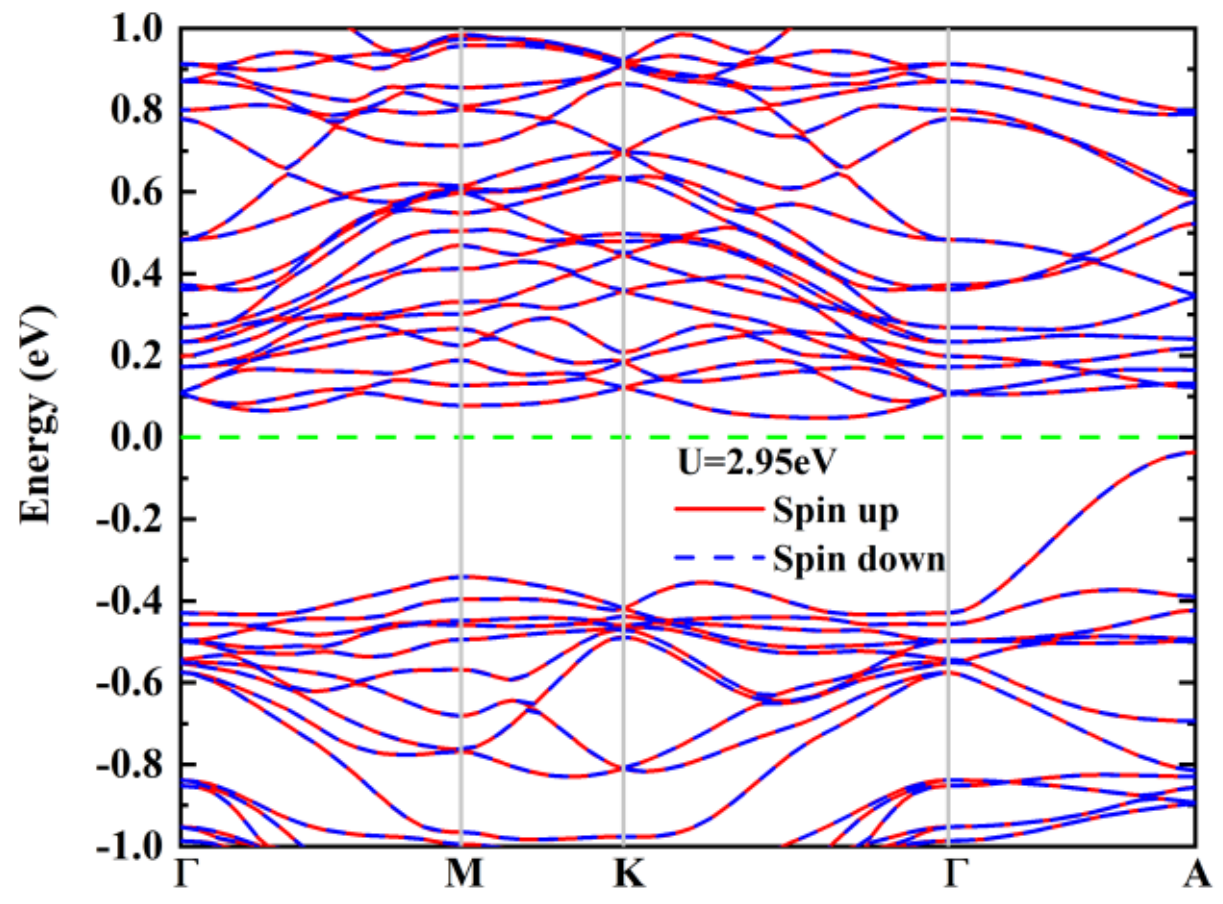

**FIG. 5.** Electronic band structure of the AFM $\sqrt{13}\times\sqrt{13}$ CCDW phase calculated using a 1×1×2 supercell within GGA+$U$.

The metal-insulator transition accompanied by the CCDW transition had been observed in the archetypal CDW system, 1$T$-$TaS_2$. However, the origins of the insulating phase and metal-insulator transition are still under extensive debate and have attracted huge interest [7,8,61]. It would be interesting to explore the possibility of realization metal-insulator transition by investigating the electronic correlation interactions and magnetic ordering in the CCDW phase of 1$T$-$NbS_2$. We first calculate the electronic structure within spin-polarized GGA for the CCDW phase. Relative to the non-spin-polarized results, the energy is only lowered by a tiny value of 0.06 meV/$NbS_2$ within spin-polarized calculations. The small energy difference implies that electronic correlation effect is not the essential reason of the distortion to the CCDW phase, which is consistent with the broad bandwidth of the Nb-4$d$ bands in **Fig. 2(a)**. Furthermore, the spin-up and spin-down subbands are identical to each other (**Fig. S3** in the Supplemental Material [59]), and magnetic moment does not appear, implying the CCDW phase is still essentially a nonmagnetic state. Even though the Coulomb correlation effect and spin-orbit coupling (SOC) interactions are taken into account, the CCDW phase remains nonmagnetic metal. As shown in **Fig. S3** of the Supplemental Material [59], although the in-plane gap has been enlarged by the on-site electron-electron interaction $U$, there is still no trace of magnetic behavior, indicating that it is hard to generate an energy splitting for the two spin channels and unavailable to realize a magnetic ordering state in the CCDW phase. In addition, SOC interactions only play minor impact on the electronic structures, so it is hard to change the out-of-plane metallic transport property and the nonmagnetic state [12,40,53].

In CCDW phase of 1$T$-$NbS_2$, each cluster is constituted with 13 Nb-atoms and 26 S-atoms, implying that there is one isolated 4$d$ electron in every Star-of-David cluster. Such an electronic configuration can create a half-filled band and result in a metallic behavior when the Coulomb correction is absent. In fact, no matter the spin-polarized, Coulomb corrections or SOC interactions has been considered or not, the insulating property or magnetic behavior does not appear basing on a 1×1×1 unit cell of the CCDW phase (**Fig. S3** of the Supplemental Material [59]). The metallic behavior of the CCDW phase solely arise from the out-of-plane dispersion, where the interlayer coupling effect delocalizes the in-plane states and suppresses the spin polarization, and finally leading to a nonmagnetic ground state [12,65].

However, if the stacking of layers doubles the cell of the CCDW phase, the insulating state can be expected by including even number of orbitals. Therefore, we construct a 1×1×2 supercell by doubling the cell of the CCDW phase along the layer stacking direction, and artificially assigned an interlayer AFM ordering by setting antiparallel magnetic moments direction in near-neighbor layers. Unfortunately, without Coulomb corrections the system does not favor AFM order and still converges to the nonmagnetic metallic state. Only AFM ordering and Coulomb corrections have been included simultaneously, a tiny band gap can be opened up (**Fig. 5** and **Fig. S4** of the Supplemental Material [59]). Similar to the monolayer 1*T*-$NbS_2$ [42], the magnetic moments almost arise from the central Nb atoms of the cluster, whereas the contributions from the edge of Star-of-David cluster are negligible.

Previously theoretical investigations show that the monolayer 1*T*-$TaS_2$ is ferromagnetic (FM) Mott insulator and the bilayer is interlayer AFM insulator in the CCDW phase, whereas the ground state of the bulk phase depends on $U_{eff}/W$ ($U_{eff}$ is the effective Coulomb correlation and W is the out-of-plane bandwidth) [12,53]. Coulomb interaction is indispensable to describe the ground state of CCDW system. For the case of 1*T*-$NbS_2$, there is one isolated 4*d* electron in every Star-of-David cluster of the CCDW phase. According to conventional band theory, odd number orbital should form a half-filled band, and result in a metallic behavior without consideration of on-site electronic correlation. As shown in **Fig. S4** of the Supplemental Material [59], the metallic behavior of the CCDW phase preserves up to a moderate Coulomb correlation of 2 eV. Along with increasing electronic correlation, Mott gap is opened up and metal-insulator transition is observed. Nevertheless, we should aware that the energy difference between the insulating AFM state and the metallic nonmagnetic state is only about 0.5 meV/$NbS_2$ ($U$ = 2.95 eV), indicating that the critical temperature for the out-of-plane AFM ordering is very low. Therefore, it would be interesting to further clarify the existence of insulating magnetic ordering state in bulk 1*T*-$NbS_2$ at low temperatures by magnetic property characterization and transport property measurement.

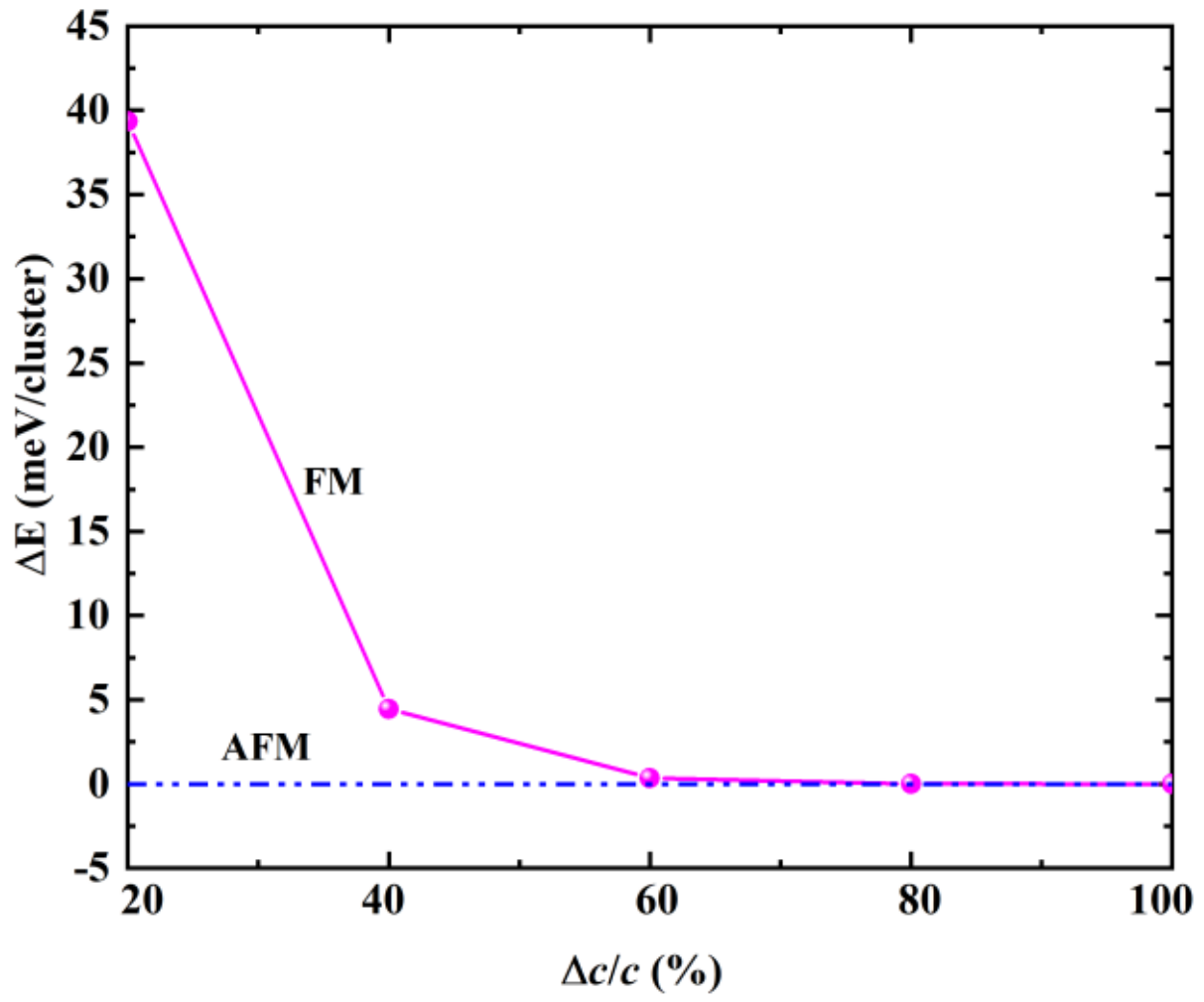

**FIG. 6.** Energy differences between FM and AFM states for the CCDW phase along with enlarging interlayer distance ($\Delta c/c$, $c$ is the lattice constant along the layer stacking direction). The calculations are carried out for the $\sqrt{13} \times \sqrt{13}$ CCDW phase within 1×1×2 supercell by GGA+$U$ ($U$ = 2.95 eV).

Moreover, when the dimensionality reduces to monolayer limit, previous study has shown that FM Mott insulating states can emerge in the CCDW phase of the 1$T$-$MX_2$, which are attributed to a consequence of the 2D geometry [12,13,40-42,65]. Recent theoretical predictions pointed out that the CCDW phase of the single-layer 1$T$-$NbS_2$ exhibits FM insulating nature [42]. However, we find a pre-set FM state converges to an interlayer AFM solution for the bulk 1$T$-$NbS_2$ with CCDW order, indicating that the AFM ordering is a more reasonable ground state. We further inspect the role of the interlayer coupling interactions and the reduced dimensionality by enlarging the interlayer distance. As shown in **Fig. 6**, the FM state is unstable with higher energy relative to the AFM state, but the energy differences between them reduce rapidly along with increasing interlayer distance. Obviously, the energies almost become the same when the interlayer distances enlarged by up to 60 percent. The interlayer coupling interactions become weaker and weaker. Finally, the system becomes essentially identical to the monolayer. Therefore, a FM insulating state is realized in the monolayer limit [42]. The manipulations of the transport properties and magnetic behavior by interlayer stacking manner, interlayer distances and intercalation deserve further study.

### D. Mechanism of CDW transition and pressure-induced superconductivity

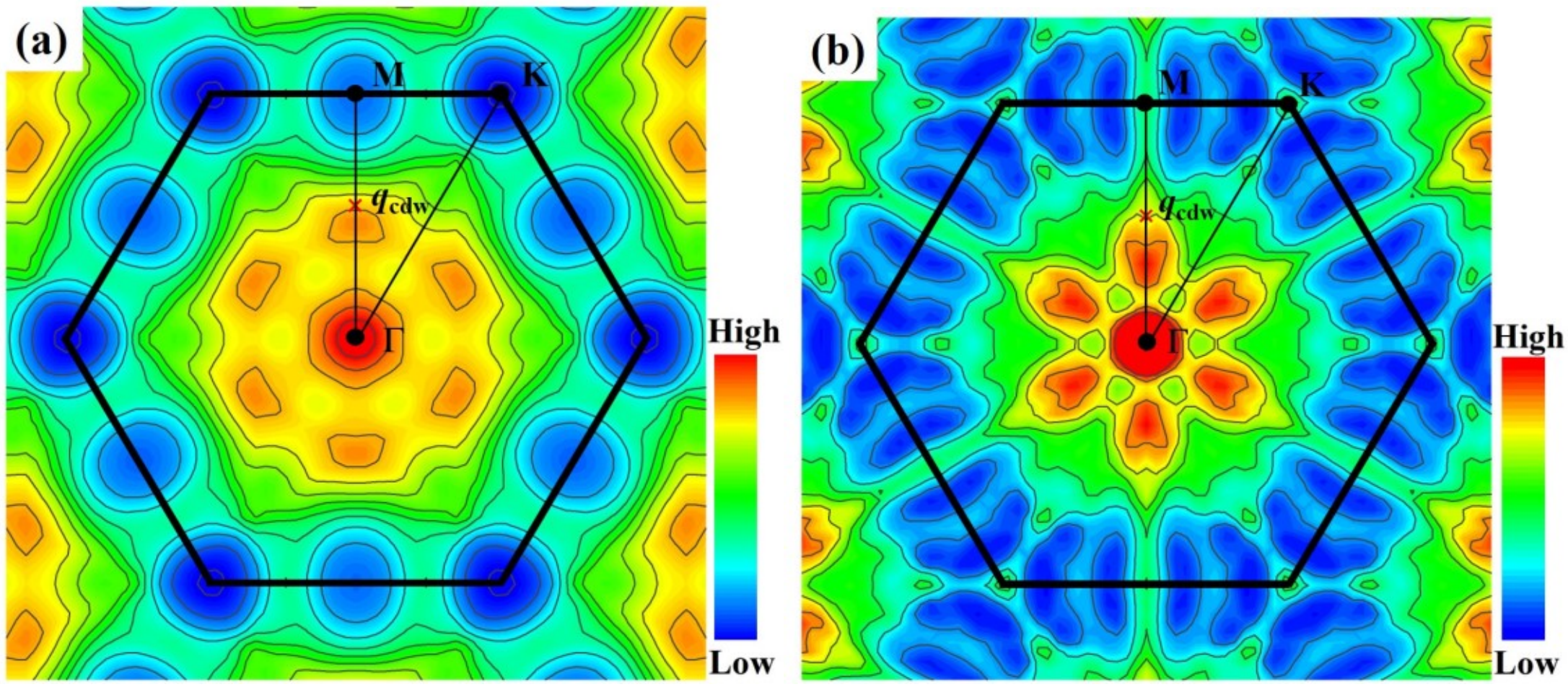

**FIG. 7.** (a) Real and (b) imaginary parts of the electron susceptibility cross section in the plane of $q_z$ = 0 for the high-symmetry bulk 1$T$-$NbS_2$, where $q_{cdw}$ is indicated by the crosses. The color bars denote the relative value.

Fermi-surface nesting has been believed as one of the driving mechanisms of CDW order

[2,3]. A quantitative evaluation of the Fermi-surface nesting can be realized by calculating electron susceptibility or Fermi surface nesting function [11]. The nesting function is the low-frequency limit of the imaginary part $Im[\chi_0(\boldsymbol{q})]$ of the bare electronic susceptibility in the constant matrix element approximation, whereas the real part $Re[\chi_0(\boldsymbol{q})]$ of the bare electronic susceptibility determines the stability of the electronic system. Therefore, if the CDW order originates from Fermi-surface nesting, peak will appear simultaneously in both of the $Im[\chi_0(\boldsymbol{q})]$ and $Re[\chi_0(\boldsymbol{q})]$ at the $\boldsymbol{q}_{\mathrm{CDW}}$ [10,11]. The real and imaginary parts of the electron susceptibility are defined as:

$$Re[\chi_0(\boldsymbol{q})] = \sum_k \frac{f(\varepsilon_k) - f(\varepsilon_k)}{\varepsilon_k - \varepsilon_{k+\boldsymbol{q}}} \tag{1}$$

$$Im[\chi_0(\boldsymbol{q})] = \sum_k \delta(\varepsilon_k - \varepsilon_F)\,\delta(\varepsilon_{k+\boldsymbol{q}} - \varepsilon_F) \tag{2}$$

where $f(\varepsilon_k)$ is the Fermi-Dirac function, $\varepsilon_k$ are the corresponding Kohn-Sham energies, and $\varepsilon_F$ is the Fermi level.

For the undistorted bulk 1*T*-$NbS_2$, a grid of 18×18×9 *k*-point is used to calculate the DFT eigenvalues for further evaluation of the electron susceptibility. Then the DFT results are used to construct a tight-binding (TB) model Hamiltonian with maximally-localized Wannier functions (MLWF) method using the WANNIER90 code [66,67]. As shown in **Fig. 2(a)**, the *d*-like bands split off from the *p*-like bands by an obvious gap around Fermi level. Therefore, the TB Hamiltonian is constructed with five Nb-*4*d orbitals. The bands derived from five-orbital model match perfectly the DFT bands of the bulk 1*T*-$NbS_2$ in the high-symmetry phase (**Fig. S5** of the Supplemental Material [59]). Subsequently, the resulting TB Hamiltonian is used to calculate the bare electron susceptibility [68-71]. Although the imaginary part of the electron susceptibility shows the maximum around **Γ** point, which is owing to intraband contributions from a weakly dispersing band and is irrelevant for the nesting [72,73]. Both of the real and imaginary parts of the susceptibility show no maxima at the positions of $q_{\mathrm{cdw}}$ (**Fig. 7**). Consequently, the Fermi-surface nesting cannot account for the CDW instability in bulk 1*T*-$NbS_2$.

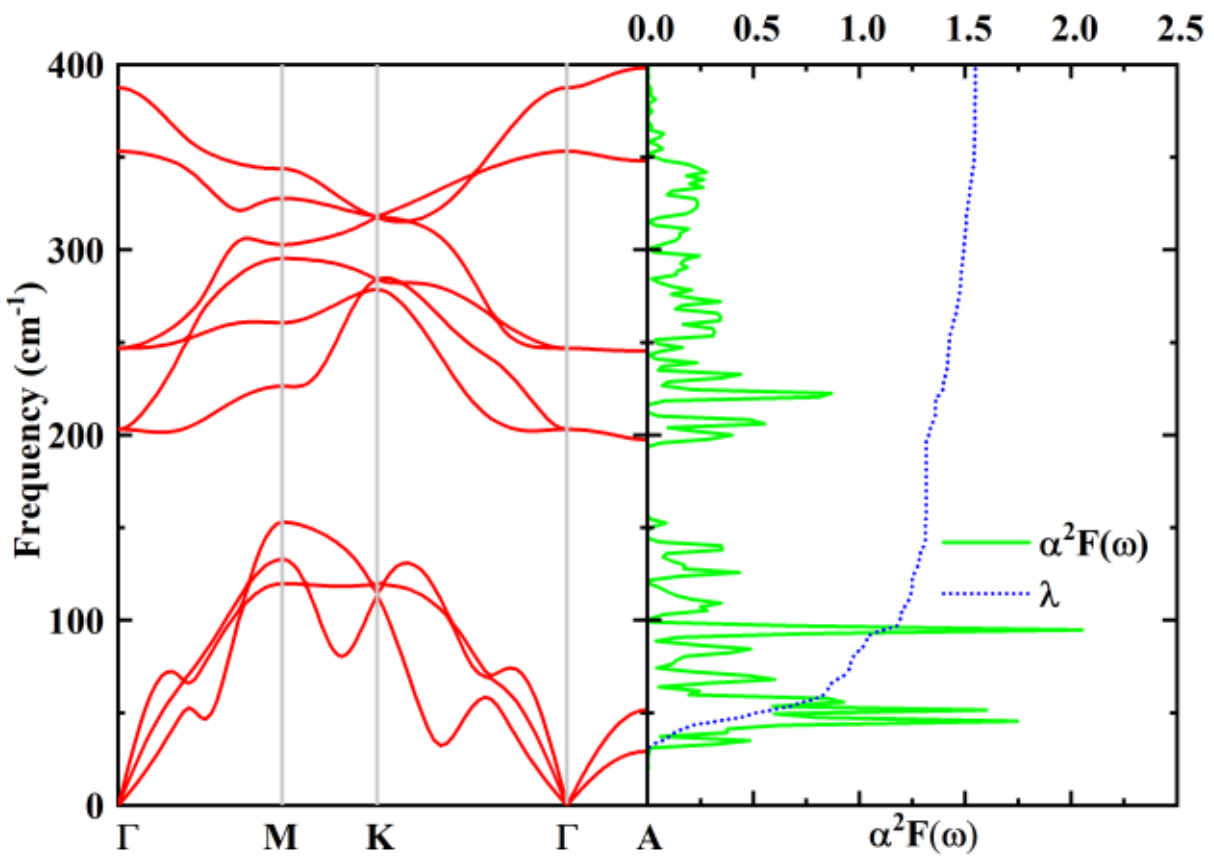

**FIG. 8.** Phonon dispersion curves and Eliashberg EPC spectral function $\alpha^2 F(\omega)$ with the integrated EPC constants $\lambda$ for the undistorted high-symmetry 1*T*-$NbS_2$ calculated within a large smearing parameter σ of 0.03 Ry.

In addition to the Fermi surface nesting mechanism, many researchers argue that the momentum-dependent electron-phonon interactions are required to explain the phonon mode softening and create CDW in quasi-2D TMDC systems [2,5,9-11]. The phonon linewidth reflects the strength of the EPC, which does not depend on the positive or imaginary nature of the phonon frequency [15,73]. Calculated phonon linewidth of the lowest phonon mode in the $q_z$ = 0 plane indicate that $q_{\mathrm{CDW}}$ is in the area with the maximum (**Fig. S6** of the Supplemental Material [59]), signifying the EPC may play an important role in CDW formation in 1*T*-$NbS_2$. Furthermore, even though it is problematic to calculate the quantitative EPC of the undistorted phase due to the imaginary frequencies around $q_{\mathrm{CDW}}$, it is reasonable to eliminate the imaginary frequencies of the phonon by increasing the electronic smearing parameter σ during the calculation procedure of the phonon dispersion [74-76]. An electronic smearing of 0.03 Ry is used to stabilize the undistorted bulk 1*T*-$NbS_2$, so that the unstable CDW-related acoustic branches can be involved in EPC calculation. The smearing parameter has been carefully verified by relaxation the crystal structure and calculating the electronic band structures of the undistorted phase. As shown in **Fig. 8** and **Fig. S7** of the Supplemental Material [59], larger σ indeed removes the imaginary phonon mode, and the CDW instability is maintained as shown by the abnormal dip phonons around $q_{\mathrm{CDW}}$, whereas the smearing parameter almost plays no role in the lattice constants and band structures [74]. The EPC constant can be calculated by

$$\lambda = 2\int_0^{\infty} \frac{\alpha^2 F(\omega)}{\omega} d\omega, \tag{3}$$

where $\alpha^2 F(\omega)$ is the Eliashberg spectral function defined as

$$\alpha^2 F(\omega) = \frac{1}{2\pi N(E_F)} \sum_q \delta(\omega - \omega_q) \frac{\gamma_q}{\hbar \omega_q}, \quad (4)$$

where $N(E_F)$ is the DOS at Fermi level and $\gamma_q$ is the phonon linewidth. Then the EPC constant is calculated to be as large as 1.55, indicating that the CDW instability of bulk 1*T*-$NbS_2$ is mainly attributed to softened phonon arising from strong EPC interactions. The indispensable role of EPC in driving the CDW transition has been recognized in 1*T* polymorph of $TaSe_2$ and $TaS_2$ [15,16].

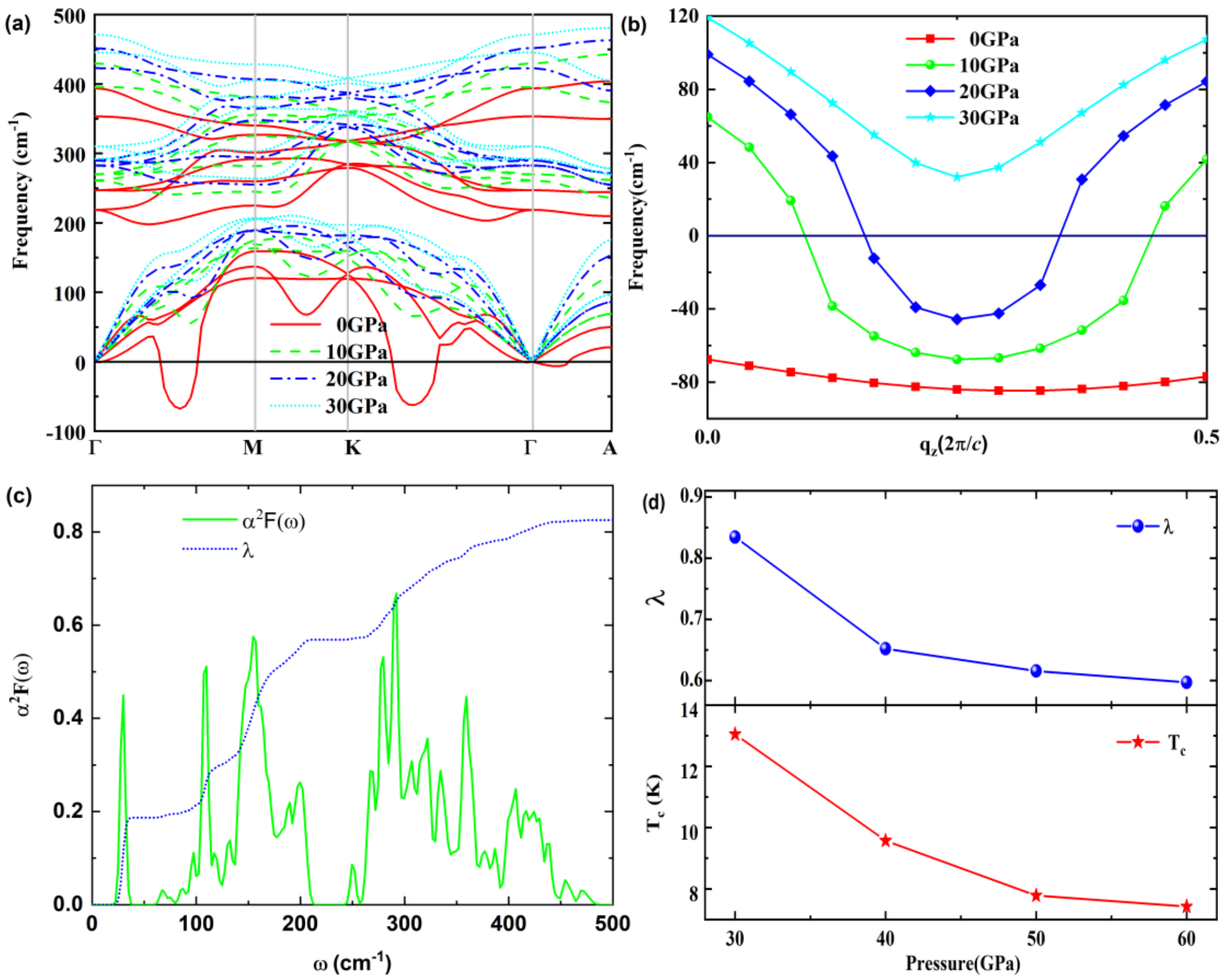

**FIG. 9.** Evolution of (a) phonon dispersion and (b) $q_z$ dependence of the unstable acoustic branch of 1*T*-$NbS_2$ under high pressure. (c) Eliashberg spectral function $\alpha^2 F(\omega)$ and the EPC integral $\lambda(\omega)$ at pressure of 30 GPa. (d) The evolutions of EPC constants ($\lambda$) and superconducting temperatures ($T_c$) of compressed 1*T*-$NbS_2$.

Previous studies show that it is possible to suppress the CDW instability upon applying pressure and trigger superconductivity in 1*T*-$MX_2$ [13,15-19,57]. As shown in **Fig. 9(a)**, the unstable modes in the bulk of 1*T*-$NbS_2$ gradually harden with increasing pressure. The unstable phonon modes of the high-symmetry structure will disappear completely and a stable phase appears above the pressure of 30 GPa [**Fig. 9(b)**]. In order to inspect the relationship between the electron-lattice interaction and the CDW instability, we have calculated the EPC constants as a function of pressure in the high-symmetry phase. Along with increasing pressures, where the 1*T* structure becomes stable, $\alpha^2F(\omega)$ shows obvious change with sharp decrease of its intensity, leading to a remarkable reduction of the EPC constant from 1.55 down to 0.82 as the CDW instability is suppressed [**Fig. 9(c)**]. Accompanied by the suppression of the CDW order, superconductivity can emerge in the bulk of 1*T*-$NbS_2$. The superconducting transition temperature $T_c$ is estimated by the Allen-Dynes-modified McMillan formula [77,78]:

$$T_c = \frac{\omega_{log}}{1.2}\exp\left(-\frac{1.04(1+\lambda)}{\lambda-\mu^*-0.62\lambda\mu^*}\right), \tag{5}$$

where the Coulomb pseudopotential $\mu^*$ is assumed to be 0.1, $\lambda$ is the EPC constant defined in **Equation (3)**, and $\omega_{log}$ is the logarithmic average frequency:

$$\omega_{log} = \exp\left(\frac{2}{\lambda}\int\frac{\log\omega}{\omega}\alpha^2F(\omega)d\omega\right), \tag{6}$$

with the Eliashberg spectral function $\alpha^2F(\omega)$ defined in **Equation (4)**.

At pressure of 30 GPa, the superconducting transition temperature $T_c$ is estimated to be 13.05 K, which is larger than those of the 1*T* polymorph of $TaS_2$ and $TaSe_2$ as well as their doping compounds $TaS_{1-x}Se_x$ [17-19]. As the CDW instability is just suppressed and the system enters into the normal phase, the original imaginary frequency becomes positive value, resulting in large $\lambda$ and high $T_C$. The EPC constants and superconducting transition temperatures of the compressed 1*T*-$NbS_2$ decrease monotonously along with the increasing pressure [**Fig. 9(d)**]. The downward trend of the superconducting transition temperatures with increasing pressure is in qualitative agreement with previous theoretically-predicted results of the 1*T* polymorph of $TaS_2$ and $TaSe_2$ [15,16]. Generally, pressure-induced lattice deformation directly causes an enhancement of $N(E_F)$ at the Fermi level and the instability of phonon vibration modes. Electronic structure calculations indicate a tiny decreases of $N(E_F)$ along with the increasing pressure, whereas the phonons of the compressed superconducting phase display an overall blue-shift compared to the ambient pressure phase 1*T*-$NbS_2$ (**Fig. S8** of the Supplemental Material [59]). Pressure compresses the lattice and render the phonon spectra gradual harden, which leads to the reduction of $\lambda$ and a consequent decrease of superconducting transition temperature.

It should be aware that superconductivity and CDW order are not absolutely exclusive with each other. Although CDW instability in pristine 1*T*-$TaS_2$ has been predicted theoretically to be suppressed by high pressure [15], the coexistence of pressure-induced superconductivity with CDW order has been observed experimentally above 4GPa [13,17-19]. The superconductivity and its coexistence with CDW order under high pressure have been explained in terms of a microscopic phase separation in real space [13]. As shown in **Fig. 9**, the phonon spectrum gradually hardens with increasing pressure. The in-plane imaginary phonon modes have disappeared above 10 GPa, but the $q_z$ dependence of the unstable acoustic branch remains negative until the pressure increases up to 30 GPa. In addition, anomalies in the acoustic branches along the **Γ-M** and **Γ-K** directions persist, signaling the incipient CDW instability of the high-symmetry 1*T*-$NbS_2$ even up to 30 GPa [15,16]. We suspect that the CDW order may coexist with the pressure-induced superconducting state till the thorough suppression of the CDW order. Further experimental studies on 1*T*-$NbS_2$ under pressure can give us the details and address this issue.

## IV. CONCLUSIONS

By employing first-principles calculation, we predict that bulk 1*T*-$NbS_2$ is unstable with softened phonon mode and will transform to stable CCDW phase with a $\sqrt{13}\times\sqrt{13}$ structural reconstruction. Electron structure calculation suggests that ODW can coexist with the CDW order in the CCDW phase. Metal-insulator transition can be realized in the CCDW phase, providing that the interlayer AFM ordering and the Coulomb correlation effect been taken into account simultaneously. Bare electron susceptibility, phonon linewidth and EPC constants calculations suggest that the EPC interactions account for the CDW instability. Furthermore, the CDW order can be suppressed by high pressure, and superconductivity emerges in the compressed structure with moderate EPC interactions. Our theoretical predictions call for further experimental study on the electronic structure and magnetic properties of the 1*T*-$NbS_2$. Moreover, it will be much interesting to study the possibility of coexistence of superconductivity and CDW order in 1*T*-$NbS_2$ phases.

## ACKNOWLEDGMENTS

The authors would like to thank Chao Cao, Wenjian Lu, Chenchao Xu, Pengfei Liu, Jianyong Chen and Yunhai Li for inspiring discussions and valuable suggestions. The work was sponsored by the National Natural Science Foundation of China (No. 11864008) and Guangxi Natural Science

Foundation (No. 2018GXNSFAA138185, 2018AD19200 and 2019GXNSFGA245006). C.A. is supported by the Foundation for Polish Science through the International Research Agendas program co-financed by the European Union within the Smart Growth Operational Programme. Z. G acknowledges the financial support from MOE tier 1 funding of NUS Faculty of Science, Singapore (grant no. R-144-000-402-114).

# Supplemental Material

## Charge density wave instability and pressure-induced superconductivity in bulk 1*T*-$NbS_2$

Wei Wang,[1] Bosen Wang,[2] Zhibin Gao,[3] Gang Tang,[4] Wen Lei,[1] Xiaojun Zheng,[1] Huan Li,[1] Xing Ming,[1*] and Carmine Autieri[5†]

*1. College of Science, Guilin University of Technology, Guilin 541004, China*

*2. Beijing National Laboratory for Condensed Matter Physics and Institute of Physics, Chinese Academy of Sciences, Beijing 100190, China*

*3. Department of Physics, National University of Singapore, Singapore 117551, Republic of Singapore*

*4. Theoretical Materials Physics, Q-MAT, CESAM, University of Liège, B-4000 Liège, Belgium*

*5. International Research Centre MagTop, Institute of Physics, Polish Academy of Sciences, Aleja Lotników 32/46, PL-02668 Warsaw, Poland*

[*] Email: mingxing@glut.edu.cn (Xing Ming)
[†] Email: autieri@magtop.ifpan.edu.pl (Carmine Autieri)

**Table S1** Atomic Wyckoff positions of the CCDW phase

| **Atom** | ***Wyckoff site*** | ***x*** | ***y*** | ***z*** |
|---|---|---|---|---|
| Nb | 1a | 0.00000 | 0.00000 | 0.00000 |
| Nb | 6g | 0.28895 | 0.07091 | 0.00079 |
| Nb | 6g | 0.63544 | 0.15267 | 0.99709 |
| S | 6g | 0.05081 | 0.17504 | 0.27403 |
| S | 6g | 0.35420 | 0.25173 | 0.27096 |
| S | 6g | 0.48565 | 0.19924 | 0.75082 |
| S | 6g | 0.97298 | 0.40846 | 0.24565 |
| S | 2d | 0.33333 | 0.66667 | 0.75405 |

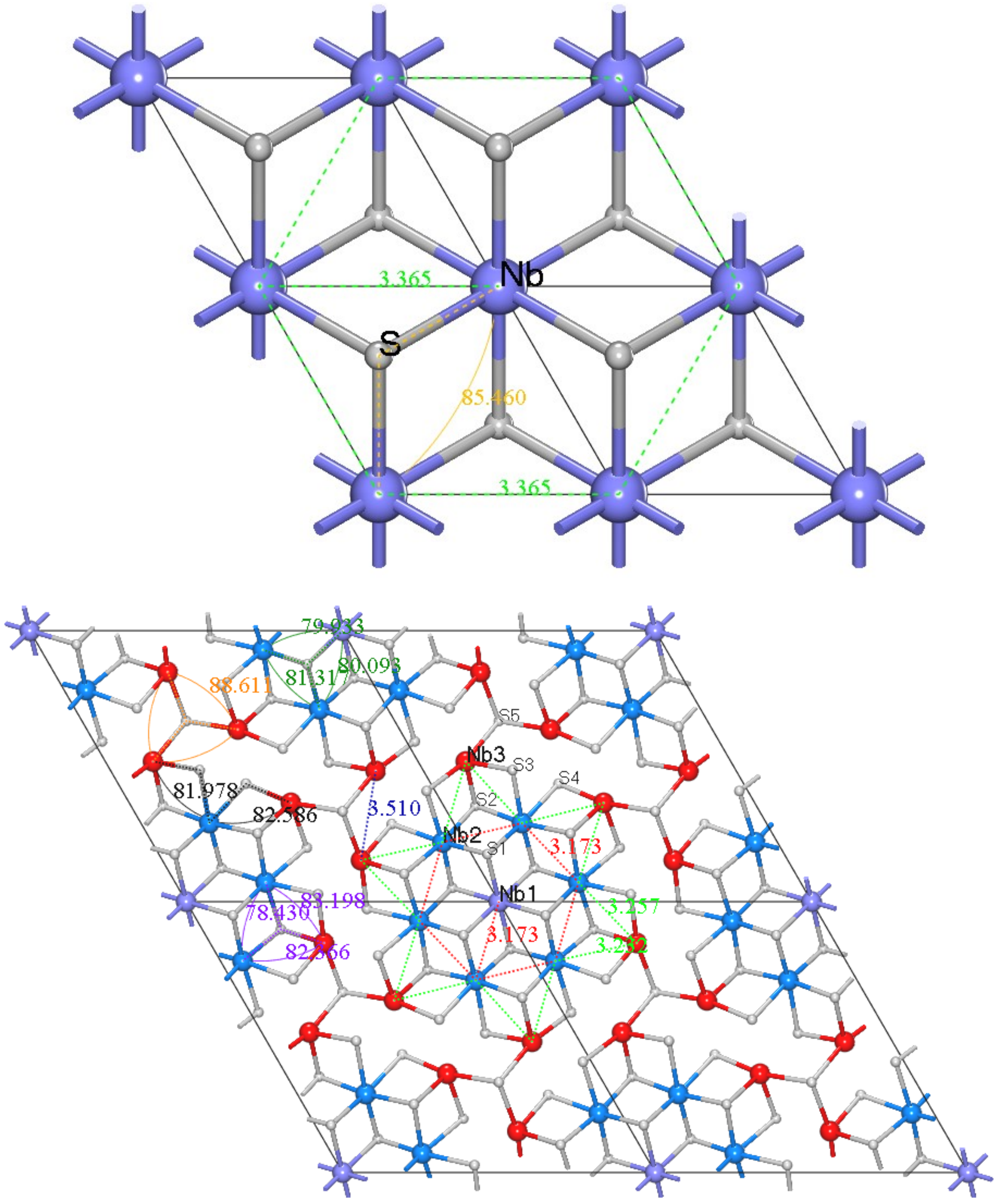

**Fig. S1** Vertical view of the optimized crystal structure for the undistorted high-symmetry phase (top) and distorted CCDW phase (bottom) of bulk 1*T*-$NbS_2$. Compared to the undistorted phase, there are three nonequivalent Nb atoms sites in the CCDW phase: The central Nb atoms in the star (purple), the peripheral Nb atoms belonging to the $\sqrt{7}\times\sqrt{7}$ cluster (blue), and more peripheral Nb atoms sites (red). These three kinds of Nb atoms form $\sqrt{13}\times\sqrt{13}$ star-of-David clusters schematized by green lines. The S atoms are shown in gray balls. The Nb-Nb distances and Nb-S-Nb bond angles are shown by the numbers.

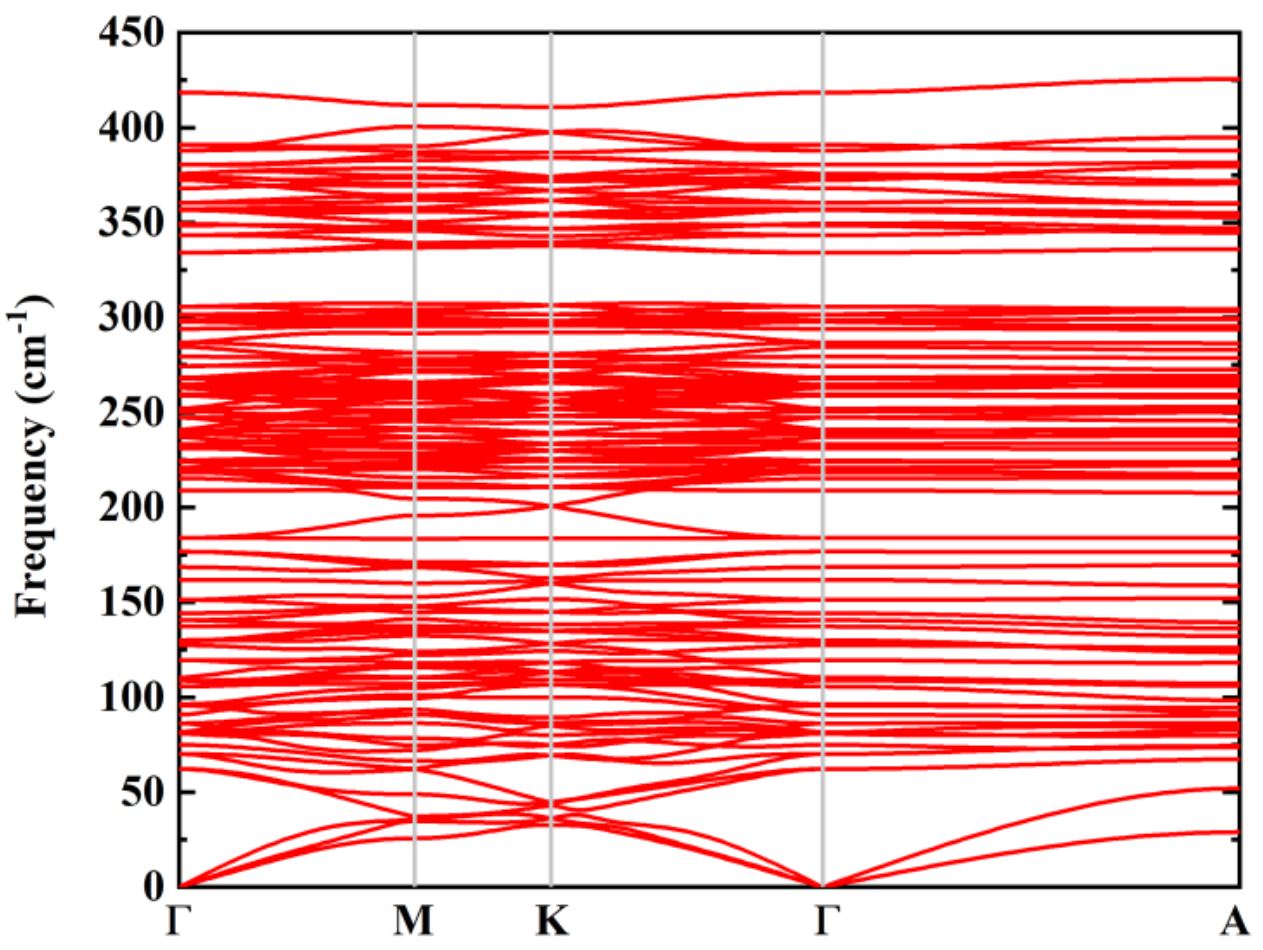

**Fig. S2** Phonon dispersion curves for the $\sqrt{13} \times \sqrt{13}$ CCDW phase of bulk 1*T*-$NbS_2$ calculated by VASP joint with Phonopy code within 1×1×2 supercell (78 atoms).

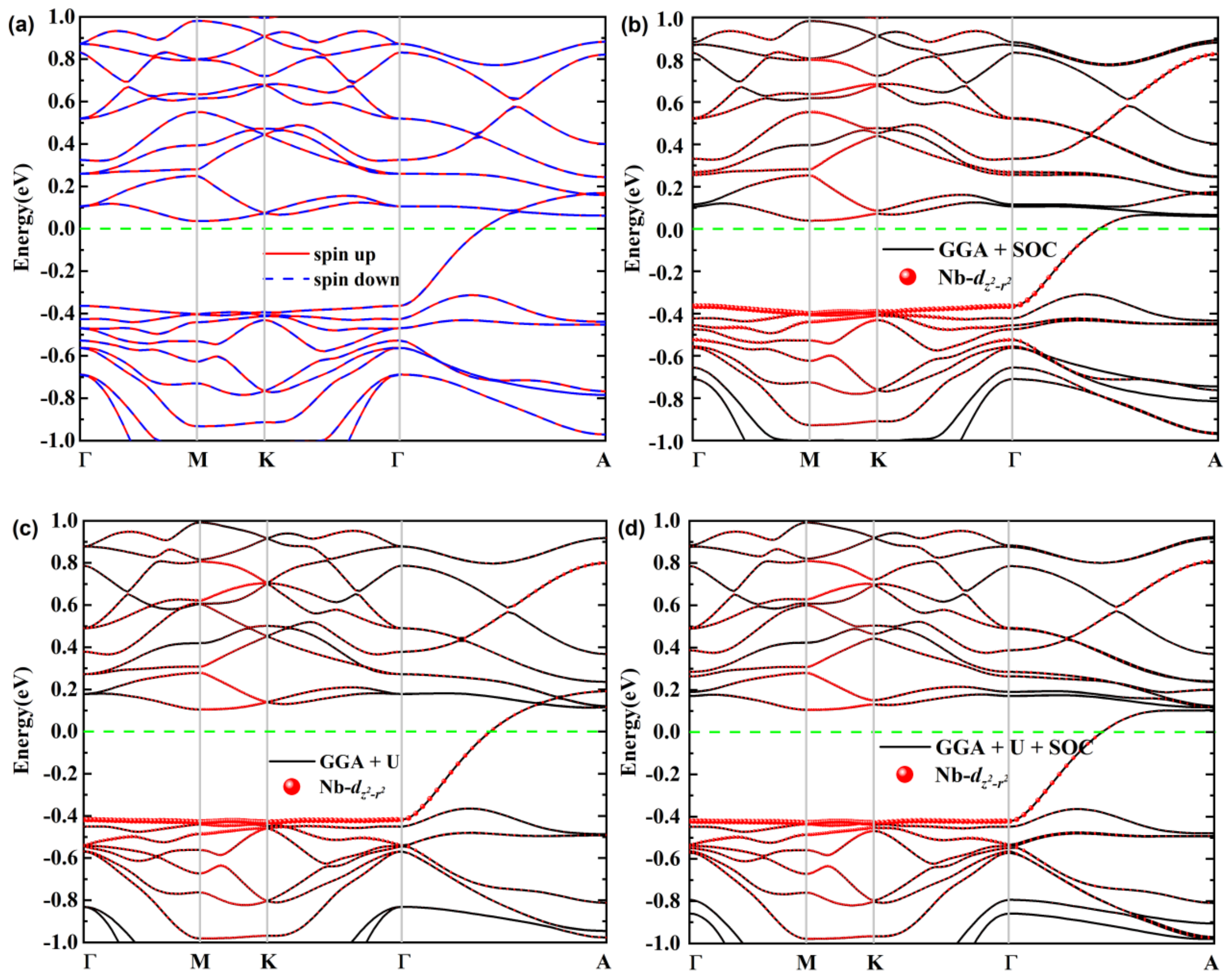

**Fig. S3** Electronic band structures of the $\sqrt{13} \times \sqrt{13}$ CCDW phase within 1×1×1 cell calculated with (a) spin polarized GGA, (b) GGA+SOC (c) GGA + *U*, and (d) GGA + *U* + SOC. The red balls represent orbital contributions with the $d_{z^2-r^2}$ character. We use *U* value of 2.95 eV here.

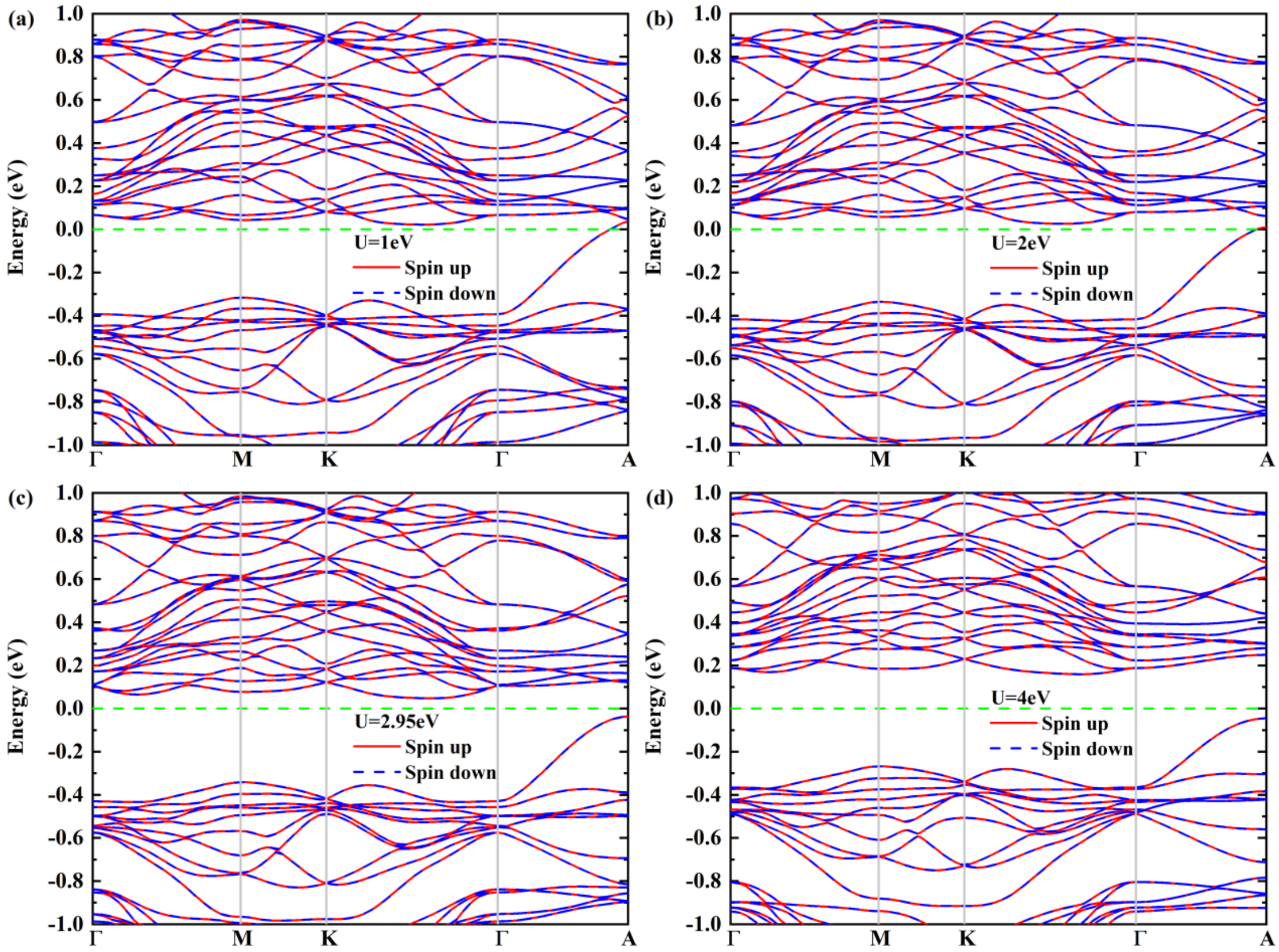

**Fig. S4** Antiferromagnetic electronic band structure of the $\sqrt{13} \times \sqrt{13}$ CCDW phase within 1×1×2 supercell calculated by GGA + $U$ with different $U$ values.

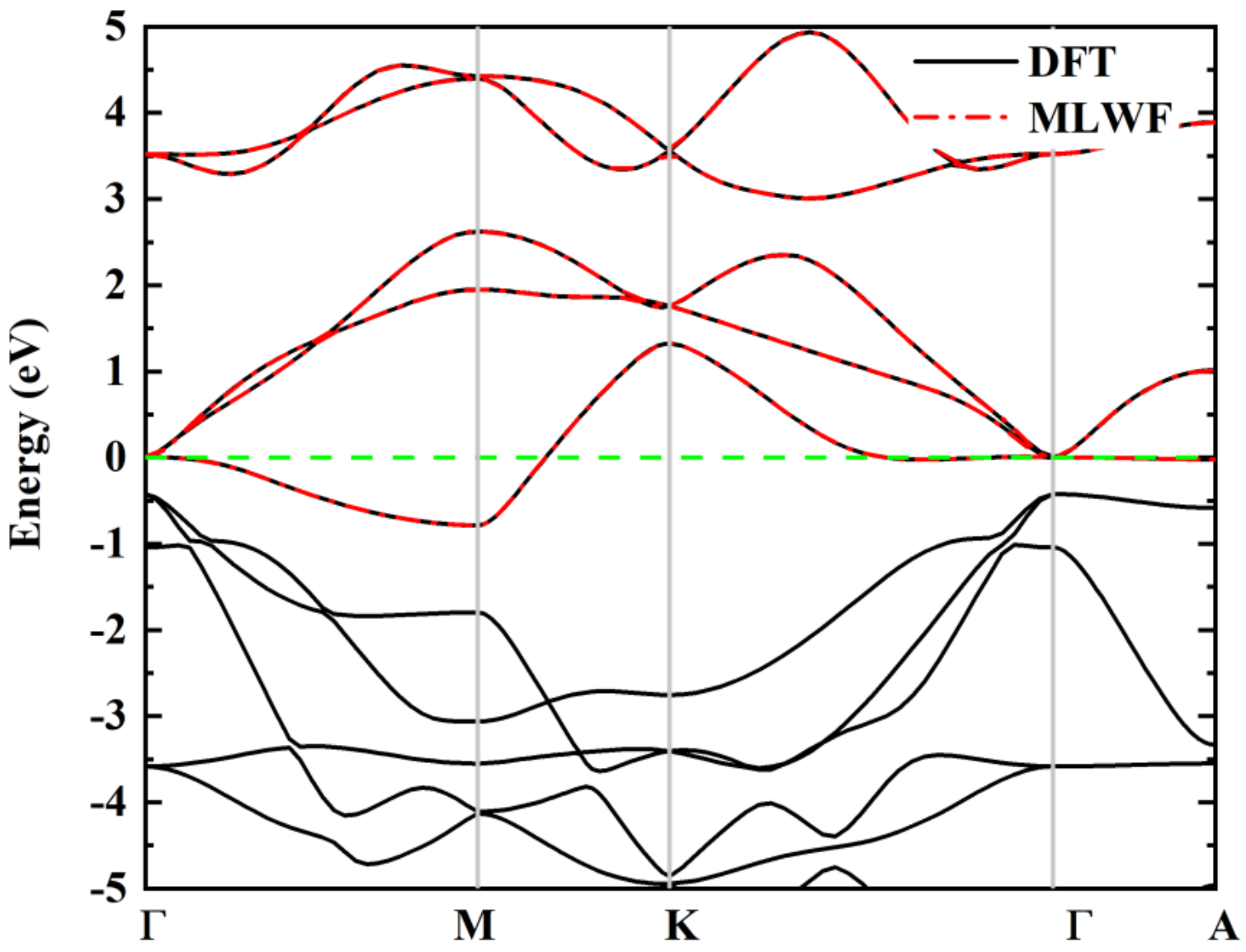

**Fig. S5** Band structure of the bulk 1*T*-$NbS_2$ in the high-symmetry phase calculated by first principles based on DFT (black solid lines) and 5-orbital *d* band model within the MLWF method (red dashed lines).

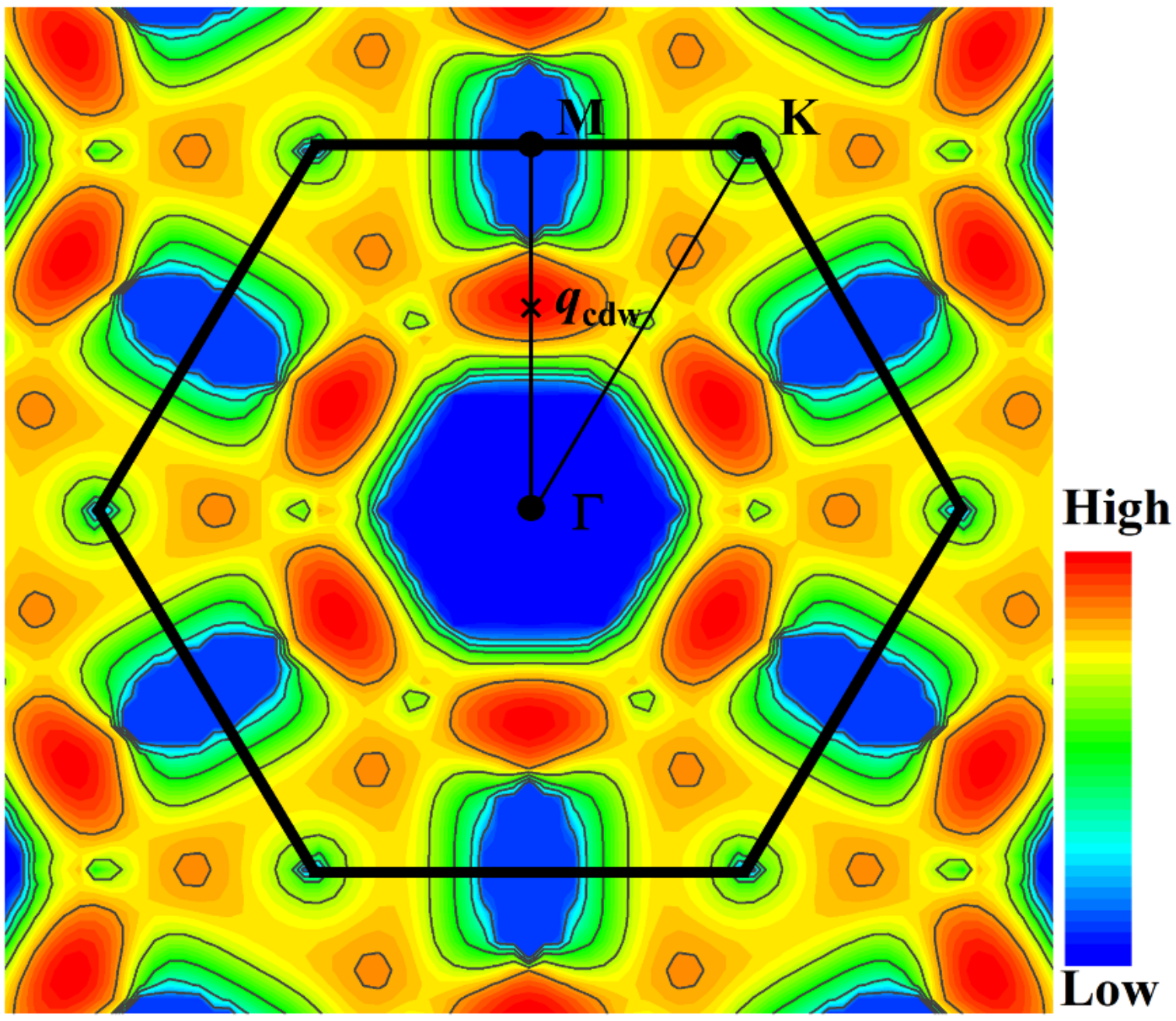

**Fig. S6** Phonon linewidth of the lowest phonon mode in the $q_z$ = 0 plane, which is calculated within a normal smearing parameter σ of 0.01 Ry for the undistorted high-symmetry 1*T*-$NbS_2$. The corresponding phonon dispersion curves have been presented in Fig. 2(c) of the main text. The color bar indicates the relative value.

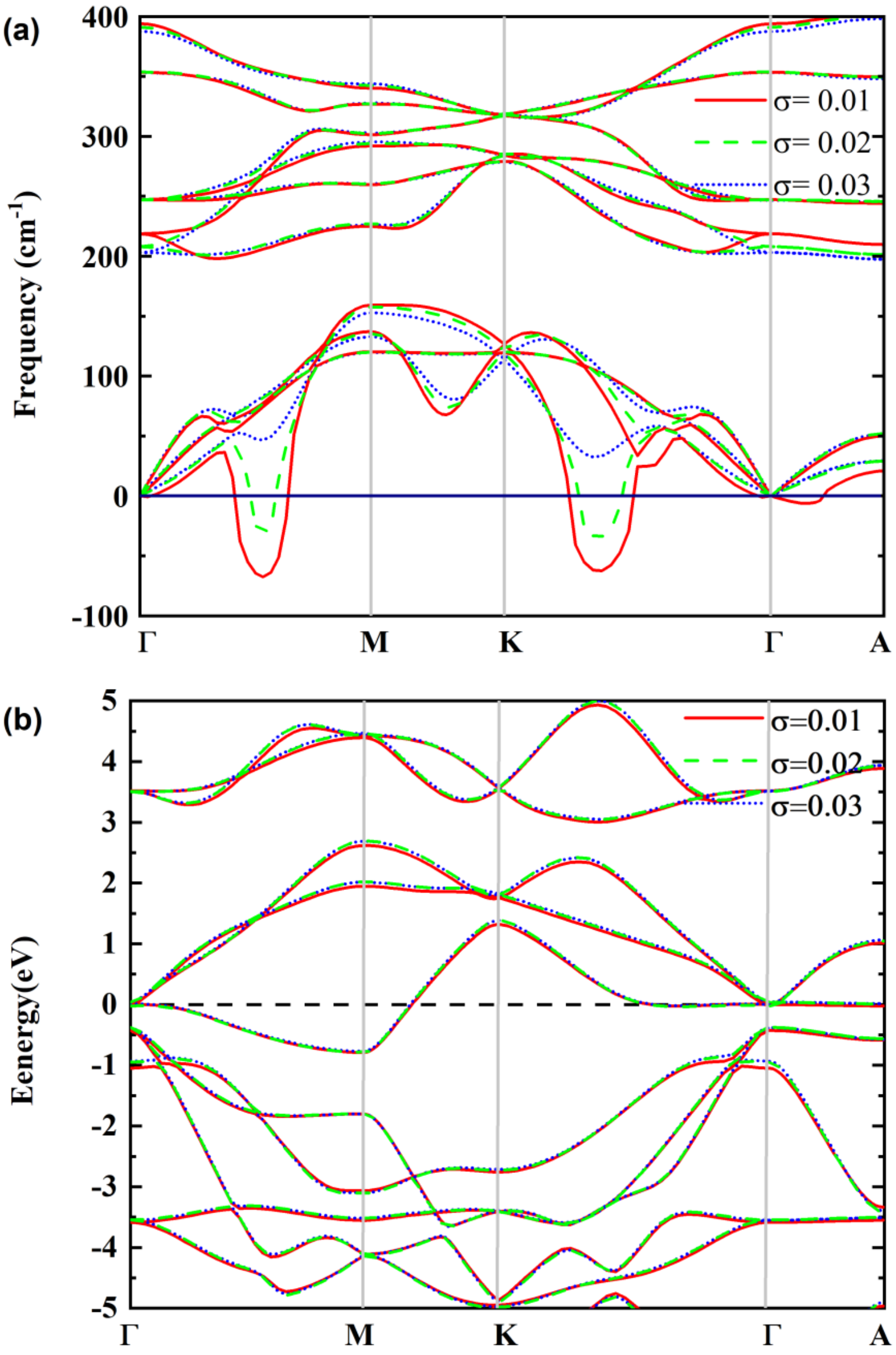

**Fig. S7** (a) Phonon dispersion curves and (b) electronic band structure of the undistorted high symmetry 1*T*-$NbS_2$ calculated within different smearing parameter σ by *QUANTUM ESPRESSO* package. The band structures are calculated with almost identical relaxed lattice constants by smearing parameter σ= 0.01, 0.02, and 0.03 Ry, where *a* = 3.365, 3.351 and 3.356 Å, and *c* = 5.954, 5.889 and 5.883 Å, respectively. Larger σ indeed removes the imaginary phonon mode, whereas the smearing parameter almost plays no role in the lattice constants and band structures.

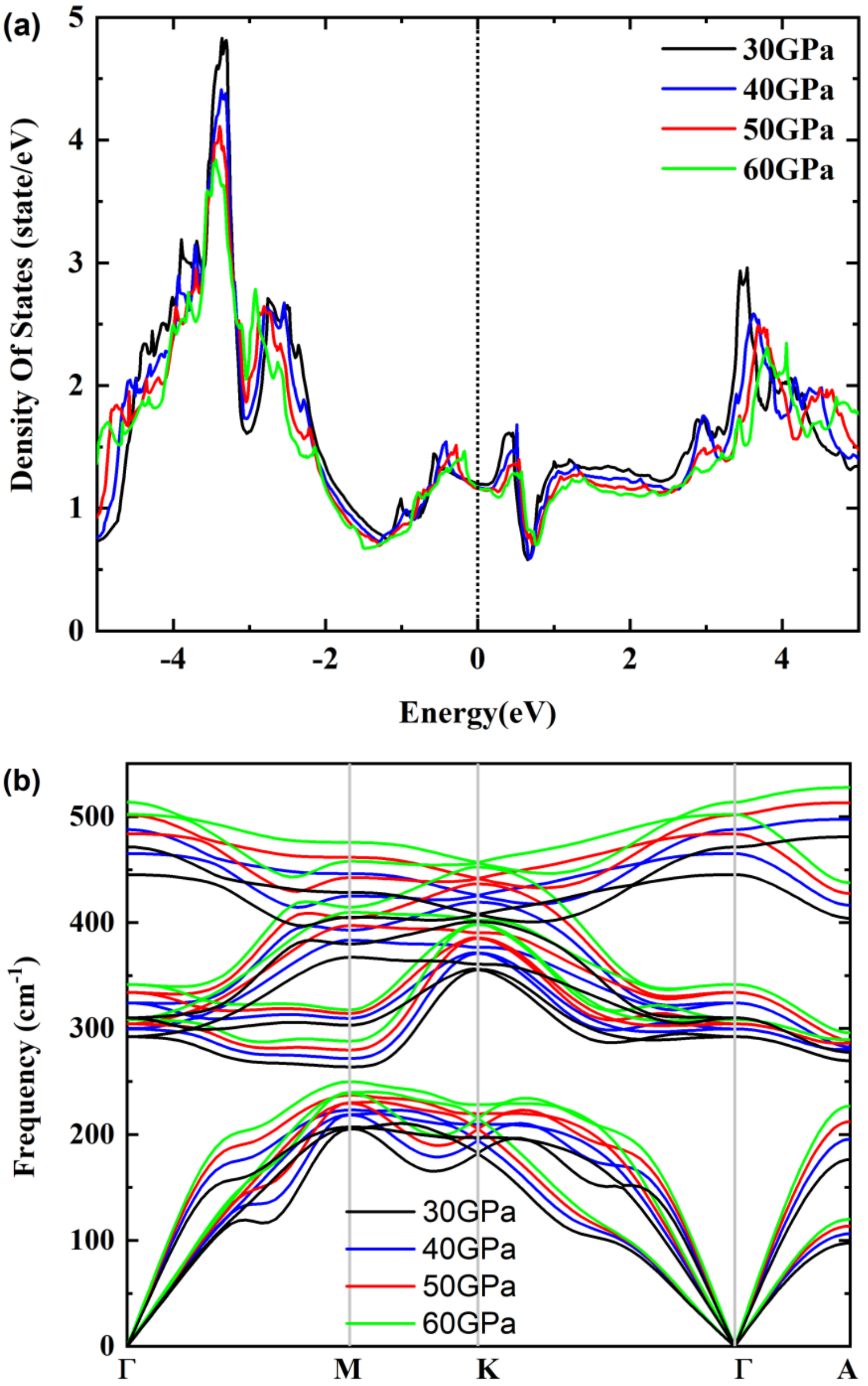

**Fig. S8** (a) Electronic density of states and (b) phonon dispersion curves for the superconducting phase of the compressed bulk 1*T*-$NbS_2$ at different pressure.